# Scenario-aware Adaptive Neuromorphic Intelligence for multimodal sensor fusion

**Kapil Bhardwaj[1], Tauseef Ejaz and Sayani Majumdar[2]**

**Faculty of Information Technology and Communication Sciences, Tampere University, 33720 Tampere, Finland**

**Email: [1]kapil.bhardwaj@tuni.fi , [2]sayani.majumdar@tuni.fi**

## *Abstract*

Multisensory Physical AI systems operating in isolated conditions must make timely and reliable decisions within strict energy and hardware budgets. In this context, the present analytical study investigates the feasibility of reliable multimodal sensor fusion using a neuromorphic approach based on memristor-based Integrate-and-Fire (I&F) neuron circuits driven by aggressively compressed representations of sensory information. Rather than relying on fixed processing and learning parameters, the proposed framework introduces scenario-dependent adaptability into individual stages of the neuromorphic decision-making pipeline, including sensor-data compression, nonlinear spike encoding, reliability assessment, and memristive learning–forgetting dynamics. By coupling multiple sensor inputs through adaptive memristive synapses, the framework investigates how reliable and correlated observations can strengthen learned associations, while weak or obsolete relationships gradually decay through nonlinear forgetting. The resulting neuronal responses are evaluated across ten representative driving scenarios derived from the *NuScenes* dataset by comparing their encoded output patterns with predefined synthetic ground-truth codes representing the expected actuator responses. The simulations demonstrate that scenario-adaptive activity-vector sizing achieves a 61.5% reduction in retained sensory information, while improving mean bit-wise agreement from approximately 55% to 61.2% compared with a fixed maximum-resolution configuration. These findings highlight the potential of introducing scenario-aware adaptability directly into the building blocks of a neuromorphic system, providing a foundation for future hardware implementations that could significantly transform the capabilities of Physical AI systems operating under stringent resource and environmental constraints.

## 1. Introduction

Rapidly advancing artificial intelligence (AI) has significantly expanded the capabilities of physical systems, including autonomous robots, unmanned aerial vehicles, and intelligent sensing platforms However, translating the algorithmic developments into practical, fully autonomous systems operating in restricted settings remains challenging, particularly because of the boundaries imposed by limited resources [1, 2]. Unlike conventional AI applications supported by highly-efficient computing infrastructures, resource-constrained Physical AI systems must continuously process sensory information and make reliable decisions in real time, often without access to cloud computing or external communication networks [3]. Such systems are therefore expected to balance computational complexity, memory availability, energy consumption, thermal constraints, and decision latency while maintaining reliable operation under changing environmental conditions. The challenge becomes more critical when incoming sensory information is heterogeneous, redundant, unlabelled, or corrupted, as the system must not only extract meaningful information but also determine which observations are sufficiently reliable to influence its learning and subsequent decisions [4].

Recent progress in edge intelligence and hardware-based on-chip learning has opened new possibilities for addressing these challenges. Among the emerging approaches, neuromorphic computing has attracted considerable attention because of its event-driven operation, distributed processing capabilities, and potential to integrate memory and computation within the same hardware architecture. In particular, spiking neural networks (SNNs), supported by emerging memory technologies such as memristive devices, offer opportunities for implementing adaptive learning and inference while reducing dependence on conventional processor–memory data transfers. Nevertheless, enabling continuous learning under limited memory and computational resources remains a major challenge, particularly when new observations must be incorporated without unnecessarily weakening previously acquired knowledge [5-7].

Several architectural-level solutions have attempted to address this learning–memory trade-off. Early developments in conventional artificial neural networks (ANNs) primarily focused on catastrophic forgetting, with Kirkpatrick et al. [8] introducing Elastic Weight Consolidation (EWC) to protect parameters important for previously learned tasks. A related strategy, Synaptic Intelligence (SI) [9], estimates parameter importance during training and regulates subsequent updates, with evaluations across Split-MNIST, Permuted-MNIST, and CIFAR benchmarks. Rather than restricting parameter modifications, Progressive Neural Networks [10] preserve previously acquired knowledge by introducing additional network columns for successive tasks, although this comes at the expense of increasing memory and computational requirements. Learning without Forgetting [11], on the other hand, employs knowledge distillation to accommodate new tasks without requiring access to the original training data. More recently, Bonnet et al. [12] introduced Metaplasticity from Synaptic Uncertainty (MESU), enabling uncertainty-dependent learning and forgetting without explicit task boundaries. Across 200 sequential Permuted-MNIST tasks, the method maintained approximately 90% accuracy on the five most recent tasks, compared with below 70% for a reported comparison approach. Alongside these advances in continual learning, improvements in multimodal perception have demonstrated the importance of efficient sensor-information representation. For instance, TransFusion [13] employed transformer-based Camera–LiDAR associations, achieving 68.90% mean average precision (mAP) and 71.68% nuScenes Detection Score (NDS), compared with 65.52% mAP and 70.23% NDS for its LiDAR-only configuration. Similarly, BEVFusion [14] combined heterogeneous sensor features within a unified bird's-eye-view representation, reporting 1.9-fold lower computational cost and more than 40-fold faster BEV pooling relative to the respective evaluated baselines.

While these ANN-based developments demonstrate progress in preserving knowledge and efficiently processing multimodal information, SNN-based architectures have explored additional possibilities for online learning through temporal spike processing and locally computed synaptic updates. Shrestha and Orchard [15] introduced SLAYER, enabling the optimization of synaptic weights and axonal delays through temporal error backpropagation, with evaluations on MNIST, N-MNIST, DVS Gesture, and TIDIGITS datasets. Moving toward more local learning mechanisms, Bellec et al. [16] developed e-prop, which combines eligibility traces with learning signals to support online adaptation in recurrent SNNs without requiring full backpropagation through time. Building on this direction, Deep Continuous Local Learning (DECOLLE) [17] demonstrated continuous local weight updates, achieving a classification error of 4.46 ± 0.16% on DvsGesture, compared with 6.36 ± 0.49% reported for SLAYER under the evaluated comparison. Further reductions in training-memory requirements were explored through Online Training Through Time (OTTT) [18], whose memory requirements remain independent of the number of simulated time steps, with demonstrations on CIFAR-10, CIFAR-100, ImageNet, and CIFAR10-DVS. Beyond computational efficiency, Skatchkovsky et al. [19] incorporated synaptic uncertainty into Bayesian continual learning for SNNs, achieving 85.44 ± 0.16% accuracy on Split-MNIST compared with 77.19 ± 0.65% for the evaluated frequentist baseline, while reporting a fivefold memory advantage over a previous continual-learning approach. An alternative direction was demonstrated through spiking neural predictive coding [20], where locally computed prediction errors facilitate continuous learning from streaming observations. Collectively, these investigations establish the potential of temporal coding, local plasticity, and uncertainty-dependent learning for reducing the computational demands of adaptive neural architectures. However, their implementation generally relies on algorithmic learning rules and network-level optimization, leaving the influence of circuit configurations and physical memory dynamics on the overall learning behavior as an important consideration.

This has motivated investigations into the fundamental computational building blocks of neuromorphic systems, particularly neurons and synapses. The Integrate-and-Fire (I&F) neuron represents one of the simplest hardware-compatible neuronal models, in which accumulated input activity produces a firing event when the membrane potential reaches a prescribed threshold. When coupled with programmable memristive elements, such neurons can support associative learning by modifying the effective synaptic conductance in response to correlated sensor activity. Moreover, the nonlinear switching and time-dependent relaxation characteristics of memristive devices offer possibilities for implementing learning and forgetting through physical memory dynamics rather than relying entirely on externally programmed update rules. Early investigations by Indiveri et al. [21] established how configurable silicon neuron circuits could reproduce integration, leakage, adaptation, and firing dynamics. Subsequently, Prezioso et al. [22] demonstrated an integrated metal-oxide memristor network capable of classifying imperfect 3 × 3 pixel patterns into three classes through programmable synaptic conductance. The significance of intrinsic memory dynamics was further demonstrated by Wang et al. [23], who reproduced short- and long-term synaptic plasticity using the relaxation characteristics of diffusive memristors. Meanwhile, a neuro-transistor-based capacitive neural network [24] explored associative learning through leaky integrate-and-fire behavior and pseudo-memcapacitive synapses. Extending these concepts toward integrated neuromorphic architectures, Duan et al. [25] demonstrated volatile $NbO_x$ memristive neurons exhibiting spatiotemporal integration, nonlinear computation, and gain modulation within a 4 × 4 memristive network. Artificial dendritic structures incorporating memristive devices [26] further demonstrated nonlinear sensory-information integration, with reported power consumption more than 1,000 times lower than a CPU and 70 times lower than an ASIC under the evaluated conditions. More recently, Sung et al. [26] demonstrated simultaneous synaptic and intrinsic plasticity within a memristive device, where synaptic potentiation increased neuronal firing frequency from 30.5 to 109.8 Hz, while a 4 × 4 device array exhibited approximately threefold faster learning acquisition through a positive-feedback mechanism. Complementing these developments, Park et al. [28] experimentally demonstrated a dynamic $TiO_x$ memristor with more than $10^6$ switching cycles, a switching speed of 10 μs, and temporal and device-to-device variability of 1.39% and 3.87%, respectively, supporting reliable leaky integrate-and-fire operation.

These studies establish that neuronal and synaptic characteristics can themselves provide valuable computational functionalities, including adaptive firing, nonlinear integration, programmable learning, and time-dependent memory relaxation. However, realizing these functionalities in practical sensor-fusion systems also depends on the nature and amount of information supplied to the neuronal circuits. Most sensing platforms continuously generate high-dimensional data, whereas event-driven neuromorphic architectures require compact spike-based representations. The preprocessing and encoding stages therefore become particularly important, as they influence not only the information retained for decision-making but also the processing activity and learning behavior of the subsequent neural hardware. The challenge is more pronounced in multimodal systems, where various modalities differ considerably in their data structures, sampling rates, information content, and susceptibility to environmental disturbances. Although aggressive filtering can reduce unnecessary information processing, excessive compression may discard features essential for reliable decisions. Conversely, retaining large amounts of redundant or corrupted information may increase processing requirements and reinforce inappropriate sensor associations during learning.

Previous investigations into efficient sensor representations and spike encoding provide important insights into this trade-off. For instance, Garg et al. [29] introduced DCT-SNN, which distributes spatial information across spike time steps using the Discrete Cosine Transform, achieving 2–14 times lower inference latency while maintaining classification accuracies of 89.94%, 68.3%, and 52.43% on CIFAR-10, CIFAR-100, and TinyImageNet, respectively. Recognizing that different portions of an event stream contribute unequally to recognition, Yao et al. [30] developed a temporal-wise attention SNN capable of suppressing less informative temporal frames, reporting an accuracy improvement of nearly 19% for gesture recognition using 60 ms observations. Alternatively, Zhou et al. [31] explored single-spike temporal coding, achieving ImageNet classification accuracy within one percentage point of an equivalent conventional deep network while improving robustness against weight quantization and noise. The importance of compact event representation was further demonstrated through Event Transformer [32], which combines sparse patch-based encoding with efficient processing to reduce computational requirements. Similarly, Dong et al. [33] employed spatiotemporal event surfaces to preserve correlations among sensory events while limiting redundant information. An Aligned Compressed Event (ACE) tensor-based approach [34] achieved 98.88% accuracy on DVS128 action recognition through compact event representation, while the more recent Spike-EVPR framework [35] suppressed temporal redundancy using SNN-compatible representations, improving Recall@1 by 7.61% and 13.20% on two event-based visual place-recognition datasets.

Taken together, these developments demonstrate that the efficiency of neuromorphic information processing is influenced by both the representation of sensory information and the underlying neuronal and synaptic dynamics. Nevertheless, these aspects have largely been investigated through separate algorithmic, architectural, or device-level approaches. In particular, **the combined influence of sensor-data compression, input reliability, non-linear synaptic dynamics, and adaptable neuronal firing behavior directly on the sensor-fusion decisions remains comparatively less explored**. This becomes especially relevant for continuously operating Physical AI systems, where the severity of the encountered situation can change rapidly and the available memory and processing resources cannot always accommodate the same amount of sensory information. A relatively simple operating condition may require only a limited representation to support an appropriate response, whereas a more complex or uncertain situation may demand additional information and greater sensitivity to temporal changes. Furthermore, not every incoming observation should necessarily influence the learning process, particularly when its reliability is questionable. These considerations suggest that adaptability should not be limited to network-level learning rules but should also extend to the individual encoding and neuronal processing parameters that determine how sensory information is represented, learned, forgotten, and ultimately translated into decisions.

Our previous investigations have progressively explored this possibility through memristor-based associative learning for multimodal sensor fusion. Initially, a compact associative learning circuit (ALC) was developed as an electrical analogue of Pavlovian conditioning, demonstrating how correlated sensor events could establish learned associations and support fault-tolerant response activation [36-37]. The framework was subsequently extended to real-world sensor-data-driven pulse representations, connecting heterogeneous sensory observations with hardware-compatible associative learning [38]. Its practical feasibility was further examined using two memristive implementations: a physical KNOWM memristor and a memristor-emulator-based circuit [39-40]. More recently, a reliability-aware learning framework was investigated to assess the confidence of incoming sensor observations before allowing them to modify the memristive state. By distinguishing the use of uncertain information for inference from its participation in learning, this approach provided a mechanism for limiting erroneous synaptic reinforcement [41].

Building upon this foundation, the present work introduces an **analytical framework for scenario-adaptive, reliability-aware neuromorphic sensor fusion based on memristive I&F neurons**. The central premise is that the processing and decision-making capability of a compact neuromorphic circuit may be improved by introducing adaptability directly into its input encoding, synaptic memory dynamics, and neuronal firing behavior, rather than relying solely on increasingly complex neural networks (Figure 1). Using ten representative driving scenarios derived from the publicly available nuScenes dataset [42], the framework evaluates sample-wise scenario complexity and its influence on information compression, nonlinear pulse generation, reliability-weighted learning, and memristive memory-state evolution. The generated output responses are compared with predefined synthetic 8-bit reference codes associated with the expected driving actions, allowing the effects of individual parameter variations to be evaluated against a fixed baseline configuration. Rather than claiming a complete autonomous hardware implementation, the analysis aims to identify the opportunities and limitations of introducing adaptability into the fundamental processing and learning stages of a memristive associative learning circuit. The findings are intended to provide a foundation for future circuit- and device-level implementations of compact, energy-efficient, and fault-tolerant neuromorphic systems capable of continuously processing sensory information and supporting local decision-making under constrained hardware resources.

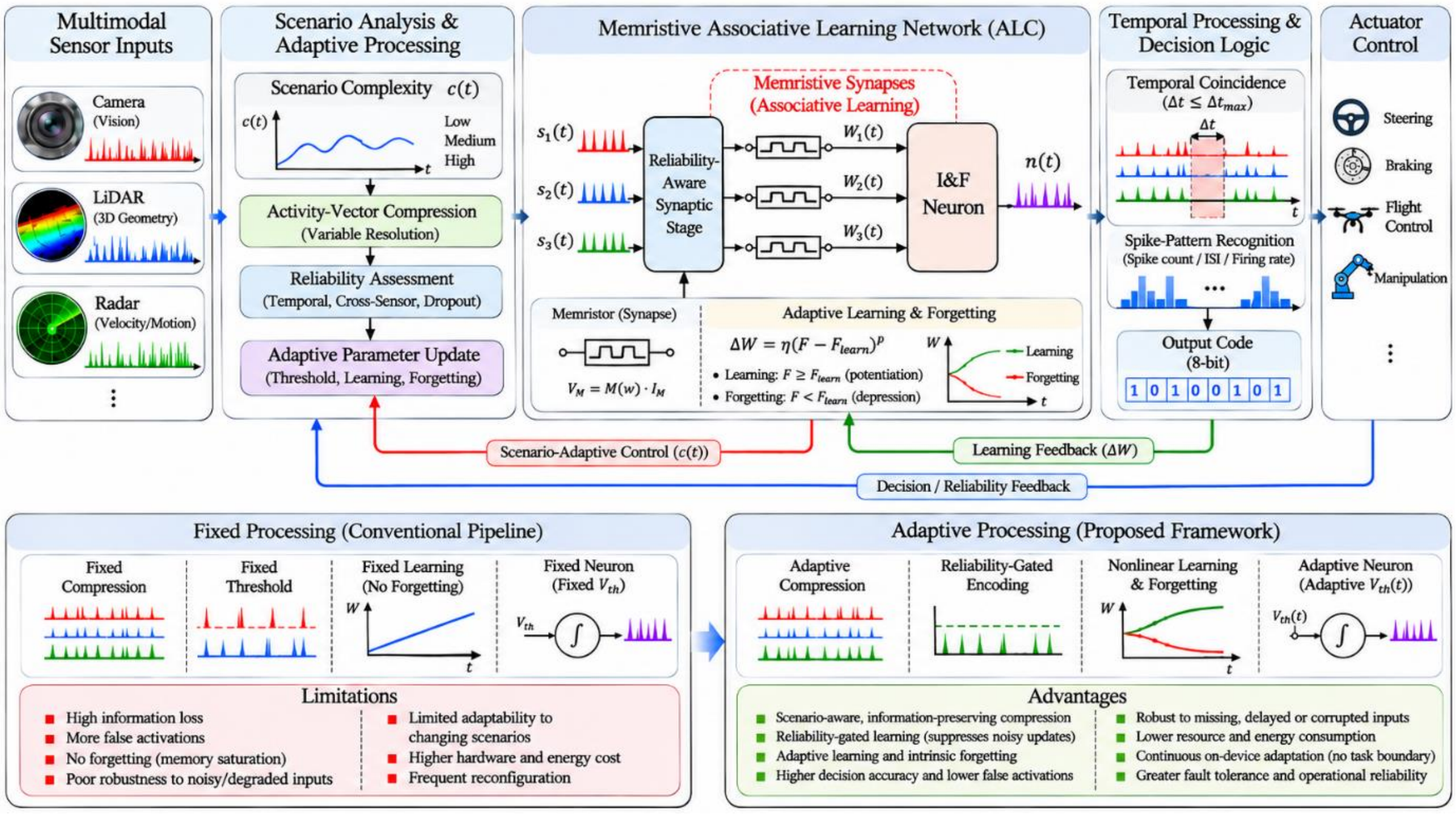

**Figure 1.** A comparison of Conventional fixed-parameter-based with the proposed scenario-adaptive neuromorphic sensor-fusion pipelines, highlighting complexity-dependent spike encoding, reliability-aware processing and adaptive memristive learning and forgetting, showing the potential of for improved decision reliability with reduced information-processing and hardware overhead.

## Models and definitions used

| No. | Parameter | Definition / Mathematical Representation |
|---|---|---|
| 1 | Scenario | A continuous sequence of synchronized observations representing a particular operating condition. |
| 2 | Sample index (k) | A discrete observation instant at which activity vectors from all sensor modalities are evaluated. |
| 3 | Activity vector ($a_s[k]$) | An N-element vector representing the spatial, temporal, or feature-level activity extracted from sensor s. |
| 4 | Activity-vector size (N) | Number of activity points retained after preprocessing. Adapted according to scenario complexity. |
| 5 | Maximum activity ($A_{max}$) | Largest activity value within a vector: $(A_{max} = \max(a_i))$ |
| 6 | Near-maximum threshold (X) | Percentage of maximum activity used to identify highly active elements: $(T_X = (X/100)A_{max})$. |
| 7 | Activity count ($C_X$) | Number of activity-vector elements satisfying $a_i \geq T_X$. |
| 8 | Event density ($D_X$) | Normalized activity count: $D_X = C_X/N$. |
| 9 | Multi-level event density | Multiple thresholds produce a compact set of densities describing the distribution of activity. |
| 10 | Scenario complexity (S) | A normalized indicator derived from temporal variation, activity variation, sensor association, and cross-sensor disagreement. |
| 11 | Association strength$(R_{ij})$ | Degree of temporal or activity correlation between sensor modalities i and j within an observation window. |
| 12 | PAM pulse ($V_P$) | Pulse-amplitude modulation maps event density into electrical pulse amplitude. |
| 13 | Nonlinear pulse encoding | Nonlinear mapping between event density and pulse amplitude: $(V_P = V_{max} f(D_X, S))$. |
| 14 | Reliability level ($A_S$) | Confidence assigned to sensor s based on temporal consistency, cross-sensor agreement, and signal quality. |
| 15 | Reliability-weighted pulse | Pulse amplitude adjusted by reliability: $(V_{p,eff} = Q_S V_p)$. Learning may also be independently gated. |
| 16 | ALC memory state ($w$) | Internal memristive state representing the learned association between correlated sensory inputs. |
| 17 | Learning | Strengthening of the memristive association following repeated, correlated, and reliable sensor events. |
| 18 | Scenario-adaptive learning | Learning dynamics adjusted according to scenario complexity and reliability. |
| 19 | Forgetting | Reduction of the stored association when reinforcement is absent, enabled by intrinsic relaxation or an implemented decay mechanism. |
| 20 | Nonlinear forgetting | State-dependent decay in which forgetting dynamics vary with association strength and device history. |
| 21 | ALC output (y[k]) | Binary HIGH/LOW response of the associative learning circuit at sample k. |
| 22 | ALC output pattern (y) | Sequence of binary responses within a decision window, decoded into an actuator command. |
| 23 | Ground truth ($y^*$) | Expected output sequence and corresponding actuator response defined for a particular scenario. |
| 24 | Decision accuracy($\mathcal{A}$) | Agreement between predicted and ground-truth patterns, evaluated using an explicitly defined comparison metric. |
| 25 | Optimal operating region Ω | Combination of encoding, reliability, neuronal, and memory parameters that achieves acceptable decision accuracy with minimal information and hardware burden. |

# 2. Theoretical Framework for Scenario-Adaptive, Reliability-Aware Memristive Associative Learning

The present simulation-based study evaluates an associative learning circuit (ALC), previously developed as an electrical analogue of Pavlovian conditioning for multimodal sensor-fusion applications [36]. The objective is to investigate whether a compact memristor-based Integrate-and-Fire (I&F) neuron architecture i.e. ALC, illustrated in Figure 2, can be adaptively configured to generate meaningful firing patterns corresponding to reliable actuator-control decisions under continuous exposure to sensor inputs from diverse operating scenarios. Central to this investigation is the nonlinear, scenario-dependent adaptability of the individual processing stages, governed by the parameterized models introduced in this section. By systematically varying these parameters, the simulations examine their individual and combined effects on neuronal firing behavior and decision accuracy.

To investigate this pathway, the framework considers three sensor modalities-Camera, LiDAR, and Radar, whose activity representations are evaluated at synchronized sample indices. The resulting firing pattern of the ALC is subsequently compared with predefined binary reference codes representing the desired actuator responses. The following subsections describe the theoretical basis of the models used in the simulation study.

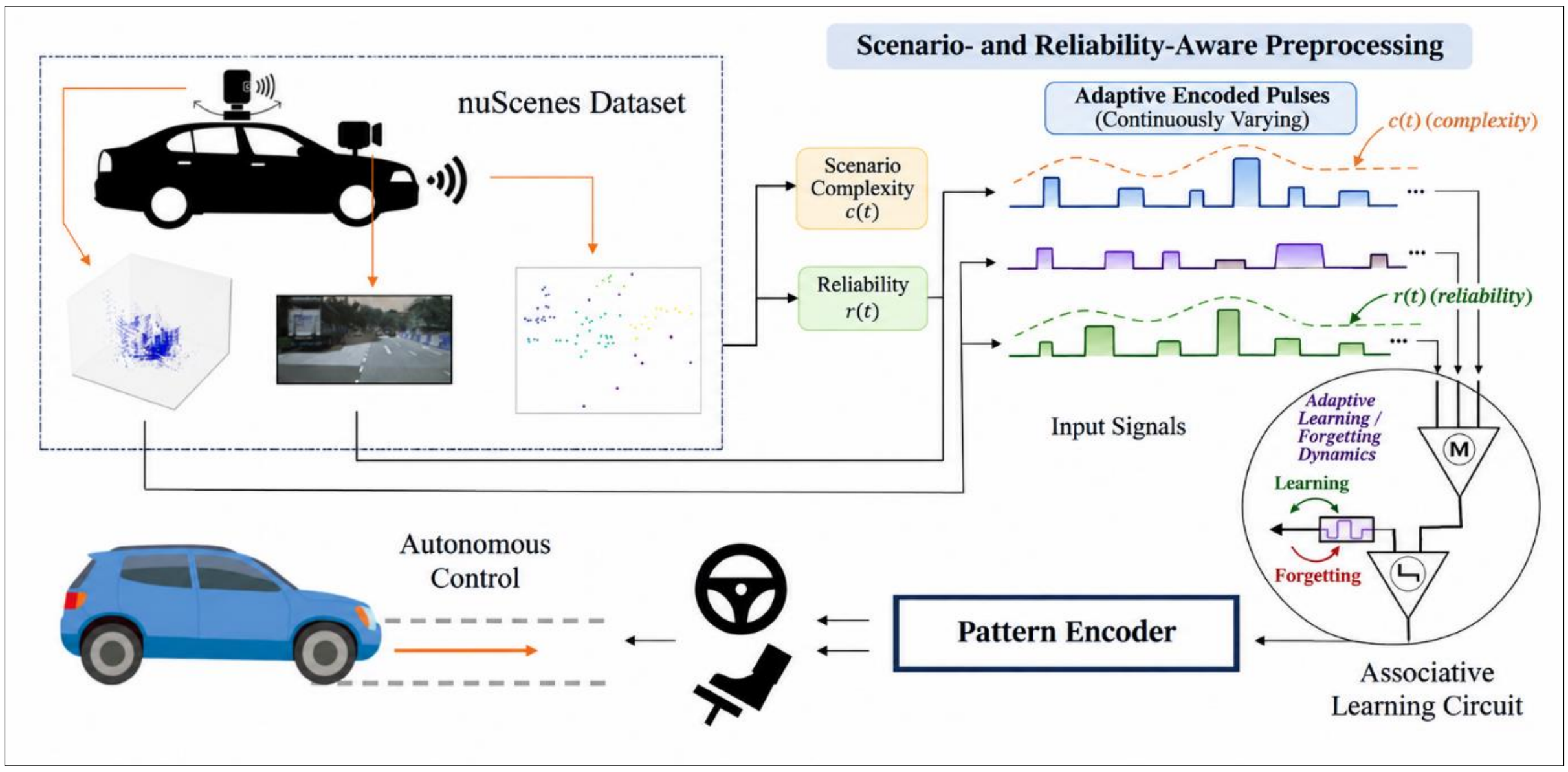

**Figure 2.** Schematic illustration of the memristor-based associative learning circuit (ALC) incorporating adaptive learning and forgetting dynamics, driven by scenario-complexity- and reliability-aware pulse encoding of multimodal sensor inputs derived from complexity-dependent event-density representations.

## 1. Multimodal sensor activity representation

The first stage involves converting the heterogeneous information obtained from the three sensor modalities into a common activity-based representation. Since Camera, LiDAR, and Radar generate fundamentally different types of information, directly processing their raw outputs within a common neuromorphic circuit would introduce substantial differences in data dimensionality and processing requirements. Therefore, each sensor is represented by an activity vector containing a selected number of activity points extracted from its corresponding observations. At a synchronized sample index k, the activity vector associated with sensor **s** is expressed as;

$$\mathbf{A_S[k] = [a_{s,1}[k], a_{s,2}[k], a_{s,3}[k] \dots a_{s,N}[k]]} \quad \mathbf{(1)}$$

At observation index k, $\mathbf{a_{S,i}[k]}$ denotes the normalized activity of the i-th retained element, and N is the number of activity points contained in the vector. The sensor index $\mathbf{s}$ represents the three considered modalities, namely $\mathbf{S_1}$= **Camera,** $\mathbf{S_2}$= **LiDAR,** $\mathbf{S_3}$= **Radar.** The activity points may represent spatial occupancy, detected-object activity, temporal changes, or other sensor-specific features relevant to the operating scenario. Although the extracted features can differ across modalities, their normalized activity representations provide a common basis for subsequent event-density computation and pulse generation. Importantly, the size of the activity vector is not fixed in the proposed framework but is adjusted according to the estimated complexity of the current operating condition.

**2. Sample-wise scenario complexity estimation**

Before determining the amount of sensory information to retain, the framework first evaluates the complexity of the encountered operating condition. Unlike conventional approaches in which a particular scenario may be assigned a fixed processing configuration, the proposed method estimates complexity independently at every synchronized sample index. This distinction is important because the severity of the operating conditions may vary considerably even within the same driving scenario.

The resulting scenario-complexity index is normalized within the range;

$$\mathbf{C_{SC} \in [0, 1]} \qquad \textbf{(2)}$$

Where $\mathbf{C_{SC}[k]}$ represents the complexity associated with the k-th synchronized observation. A value approaching zero indicates a relatively simple and stable operating condition, whereas a value approaching one represents a more complex, rapidly changing, or uncertain situation.

The complexity is estimated by combining four indicators:

$$\mathbf{C_{SC}[k] = w_1 V_A[k] + w_2 T_A[k] + w_3 S_D[k] + w_4 R_U[k]} \qquad \textbf{(3)}$$

Where, $\mathbf{V_A[k]}$represents the variation in sensor activity, $\mathbf{T_A[k]}$ describes temporal variation between successive observations, $\mathbf{S_D[k]}$ quantifies disagreement between sensor modalities, and $\mathbf{R_U[k]}$ represents the uncertainty associated with the observed sensory information. The coefficients $\mathbf{w_1}$, $\mathbf{w_2}$ , $\mathbf{w_3}$ and $\mathbf{w_4}$ determine the relative contribution of each indicator to the overall complexity. Activity variation reflects the extent to which the distribution of sensor-derived activity changes, while temporal variation captures the evolution of observations over time. Cross-sensor disagreement indicates inconsistencies between modalities observing the same environment, and uncertainty accounts for ambiguous or potentially unreliable observations. Together, these indicators provide a broader characterization of scenario severity than any individual sensor-activity measurement.

Based on the resulting complexity, three operating regions are considered:

| Complexity range | Classification | Processing Interpretation |
|---|---|---|
| $0 < C_{SC} < 0.3$ | Low | Relatively simple conditions requiring limited sensory representation |
| $0.3 \leq C_{SC} < 0.7$ | Moderate | Changing conditions requiring additional activity information |
| $C_{SC} \geq 0.7$ | High | Complex or uncertain conditions requiring greater information retention |

The estimated complexity subsequently controls the activity-vector size, event-density thresholds, and number of event-density levels. In this way, complexity estimation provides the common reference for adapting the subsequent processing stages.

**3. Scenario-adaptive activity-vector resolution**

Once the scenario complexity has been established, the number of activity points retained from each sensor is adjusted accordingly. The motivation is that complex or potentially critical situations may require a richer representation of the surrounding environment, while simpler conditions can often be described using fewer relevant features. The activity-vector size is selected as;

$$\mathbf{N[k]} = \begin{cases} \mathbf{8}, & \mathbf{C_{SC} < 0.3} \\ \mathbf{16\ or\ 32}, & \mathbf{0.3 \le C_{SC} < 0.7} \\ \mathbf{64}, & \mathbf{C_{SC} \ge 0.7} \end{cases} \qquad \textbf{(4)}$$

Where $\mathbf{N[k]}$ represents the number of retained activity points at sample index k. The investigated representative configurations use N=8, N=16, and N=64 for low, moderate, and high complexity, respectively, while an additional intermediate configuration with N=32 is considered in the parameter-comparison analysis. This adaptation allows a larger number of activity points to participate in subsequent processing when the complexity increases. Conversely, reducing N during simple operating conditions decreases the amount of sensory information that must be represented and processed.

However, reducing the vector size is meaningful only if the retained points continue to represent the relevant environmental features. Thus, the proposed approach does not assume that less information is always preferable; rather, it investigates whether the amount of retained information can be matched to the complexity of the encountered situation without substantially compromising agreement with the expected output patterns.

**4. Adaptive near-maximum activity threshold and event-density computation**

Following activity-vector selection, the next objective is to identify the highly active elements within each vector. For this purpose, a near-maximum activity threshold **X** is introduced, which determines the fraction of the maximum activity used to classify individual points as active.

Rather than assigning a fixed threshold across all operating conditions, the proposed framework considers its adaptation according to the statistical characteristics of the activity vector and the estimated scenario complexity:

$$\mathbf{X_s[k] = f\left(\sigma_{A,s}[k], \overline{A_s[k]}, C_{SC}[k]\right)} \qquad \textbf{(5)}$$

Where X[k] is the selected near-maximum activity threshold for sensor s, $\sigma_{A,s}[k]$ represents the standard deviation of its activity values, $\overline{A_s[k]}$ is the corresponding mean activity, and $C_{SC}[k]$ is the current scenario complexity. The function **f** describes the adaptive threshold-selection rule. The threshold is selected to decrease as scenario complexity increases, subject to the influence of the corresponding sensor-activity statistics. Under high-complexity conditions, a lower near-maximum threshold permits a broader range of activity values to contribute to the event-density representation. In contrast, a higher threshold during relatively simple conditions restricts the representation to the more strongly active points. Since the activity distributions of Camera, LiDAR, and Radar may differ substantially, the threshold can also vary between sensor modalities. For an activity vector with maximum value $\mathbf{A_{s,max}[k]}$, the number of highly active points can be expressed as;

$$\mathbf{C_{s,X}[k] = \sum_{i=1}^{N[k]} 1\left(a_{s,i}[k] > X_S[k] A_{s,max}[k]\right)} \qquad \textbf{(6)}$$

Where $\mathbf{C_{s,X}[k]}$ is the activity count, $\mathbf{A_{s,max}[k]}$ denotes the maximum activity within the vector, and 1(.) is an indicator function that returns one when the selected activity satisfies the threshold criterion and zero otherwise. Here, $\mathbf{X_S[k]}$ is represented as a fraction between zero and one.

The corresponding normalized event density is;

$$\mathbf{D_{s,X}[k]} = \frac{\mathbf{C_{s,X}[k]}}{\mathbf{N[k]}} \qquad \mathbf{(7)}$$

where $\mathbf{D_{s,X}[k]}$ represents the fraction of retained activity points exceeding the selected threshold. This normalization allows activity vectors of different sizes to be compared through a common density-based measure, that will ultimately provide us a shared spike encoding approach for all the modalities.

***Multilevel event-density representation***

While a single threshold provides a compact measure of highly active points, it does not necessarily capture the distribution of weaker and stronger activity within the vector. Therefore, the framework also investigates the possibility of using multiple near-maximum thresholds to generate several event-density levels from the same activity vector. Three representative threshold levels are considered:

$$\mathbf{D_{60}, D_{75}\ \text{and}\ D_{90}} \qquad \mathbf{(8)}$$

Where $D_{60}$, $D_{75}$ and $D_{90}$ represent the fractions of activity points exceeding 60%, 75%, and 90% of the maximum activity, respectively.

$$\mathbf{D_S[k] = [D_{60}[k], D_{75}[k], D_{90}[k]]} \qquad \mathbf{(9)}$$

where $D_S[k]$ contains three normalized measures of the activity distribution at the current sample index.

The number of levels used during processing is also made scenario-dependent. Low-complexity observations can be represented using a single density level, moderate-complexity observations can employ two levels, and high-complexity observations can retain three levels. For subsequent pulse generation, the selected event-density levels are combined into a **continuous composite event-density** value $\mathbf{E_S[k]}$. This value represents the retained activity distribution in a compact scalar form as a function of sampling index and provides the input to the nonlinear pulse-encoding stage. The composite density therefore depends on both the selected near-maximum thresholds and the number of event-density levels, which are themselves influenced by scenario complexity.

## 5. Nonlinear pulse-amplitude encoding

After determining the continuous composite event density, the next stage involves converting this representation into an electrical pulse suitable for driving the memristive associative learning circuit. In the proposed framework, pulse-amplitude modulation (PAM) is considered, where each synchronized sensor observation produces a pulse whose amplitude reflects the retained sensory activity. Unlike a linear encoding strategy, where pulse amplitude changes proportionally with event density, the proposed nonlinear encoding allows the amplitude to depend on both the magnitude of the composite density and its temporal variations. This enables the pulse representation to respond not only to the amount of active sensory information but also to sudden changes in the observed activity. The nonlinear PAM amplitude is described by;

$$\mathbf{P_S[k] = A_S E_S[k]^{\gamma}\left(1 + \alpha|\Delta E_S[k]| + \beta\left|\Delta^2 E_S\right|\right)} \qquad \mathbf{(10)}$$

where $\mathbf{P_S[k]}$ represents the generated PAM amplitude for sensor s at sample index k, $A_S$ is the sensor-specific amplitude scaling factor, and $E_S[k]$ represents the continuous composite event density. The exponent $\gamma$ controls the nonlinearity of the density-to-amplitude relationship, while $\alpha$ and $\beta$ determine the contributions of the first-order and second-order temporal variations, respectively. The first-order difference $\Delta E_S[k]$ represents the change in composite event density between consecutive observations, whereas $\Delta^2 E_S$ the second-order difference describes the change in that variation.

Accordingly, the first-order term responds to increasing or decreasing activity, while the second-order term allows additional emphasis on abrupt changes in the activity trajectory.

The absolute values ensure that both positive and negative activity changes can contribute to pulse amplification. The influence of the nonlinear exponent is particularly important: when $\gamma > 1$ smaller normalized event densities are relatively suppressed compared with larger values, whereas an exponent below one produces a comparatively stronger response to weaker activity levels. This behavior is particularly relevant to multimodal sensor fusion because different sensor modalities may exhibit distinct activity patterns even when observing the same environmental situation. The nonlinear encoding stage allows these variations to be reflected in the electrical pulse amplitudes before the inputs reach the associative learning circuit. Importantly, the generated amplitudes in the analytical study represent normalized or model-scaled signals. Their conversion into physical driving voltages would require calibration according to the switching characteristics and operating limits of the selected memristive device.

## 6. Reliability-aware pulse attenuation

Although the nonlinear encoding stage provides a compact representation of sensory activity, the generated pulse amplitude alone does not indicate whether the corresponding observation is reliable. A large pulse can result from either meaningful environmental activity or a corrupted sensor measurement. Therefore, the proposed framework introduces an additional reliability-assessment stage to regulate the influence of incoming sensor pulses. The reliability of sensor **s** is estimated using;

$$\mathbf{R_s[k] = Q_S[k]\left(c_0 + \alpha_R A_{agree,s}[k] + \beta_R\, T_{cons,s}[k]\right)} \tag{11}$$

subject to

$$\mathbf{0 \le R_S \le 1}$$

Where $R_S[k]$ represents the estimated sensor reliability, $Q_S[k]$ denotes the intrinsic or externally estimated quality of the sensor observation, $\mathbf{A_{agree,S}[k]}$ represents its agreement with other sensor modalities, and $T_{cons,s}[k]$ describes its temporal consistency. The coefficients $c_0$, $\alpha_R$ and $\beta_R$ determine the contributions of the baseline quality, cross-sensor agreement, and temporal consistency, respectively. The resulting reliability estimate is used to modify the generated PAM amplitude:

$$\mathbf{P_{S,eff}[k] = P_S[k] R_S[k]^{mP}} \tag{12}$$

where $P_{S,eff}[k]$ represents the reliability-weighted pulse amplitude and $\mathbf{m_P}$ determines the sensitivity of pulse attenuation to the estimated reliability. For the linear reliability-weighting case investigated in the representative pulse-attenuation results, $\mathbf{m_P=1}$. Consequently, an observation with high confidence undergoes relatively little attenuation, whereas a low-confidence observation contributes a smaller pulse amplitude to the subsequent processing stage. This mechanism is particularly important when different sensors exhibit substantially different levels of reliability.

## 7. Reliability-aware memristive associative learning

The central objective of the proposed framework is to regulate the learning behavior of the memristive element incorporated within the I&F neuron circuit. In a memristive associative learning architecture, correlated sensor-derived pulses modify the internal memory state, thereby strengthening the association between sensor modalities. These learned associations subsequently influence the effective neuronal drive and, consequently, the generated firing patterns. However, directly updating the memristive state in response to every correlated input may produce inappropriate learning when one or more observations are unreliable. To address this limitation, the effective learning strength is made dependent on the estimated confidence of the sensor association. The reliability-dependent learning coefficient is expressed as;

$$\boldsymbol{\eta}_{\mathbf{eff,ij}}[\mathbf{k}] = \boldsymbol{\eta}_{\mathbf{0}}\mathbf{R}_{\mathbf{ij}}[\mathbf{k}]^{\mathbf{m}} \tag{13}$$

where $\eta_{\mathrm{eff,ij}}[\mathrm{k}]$ represents the effective learning coefficient associated with sensor pair (i,j), η is the nominal learning coefficient under full confidence, $\mathrm{R_{ij}}[\mathrm{k}]$ is the confidence associated with the corresponding sensor pair, and **m** determines the nonlinear sensitivity of learning to reliability. When $\mathrm{R_{ij}}[\mathrm{k}]$ approaches one, the effective learning coefficient approaches its nominal value, allowing the correlated observations to reinforce the memristive association. Conversely, when the pairwise confidence becomes small, the effective learning coefficient decreases, thereby limiting the amount of reinforcement produced by uncertain observations. This reliability-dependent mechanism provides control over learning at the level of sensor pairs rather than considering each modality independently. For instance, even when two sensors produce correlated activity, their association can receive limited reinforcement if the corresponding observations have low confidence. The significance of this approach is that the circuit can continue to process uncertain information for inference while restricting its influence on the stored associations. In this way, the learning process becomes selective with respect to the evidence that is allowed to modify the memristive state.

***Minimum reinforcement requirement for associative learning***

In addition to reliability weighting, the learning process incorporates a minimum reinforcement threshold to prevent weak or coincidental sensor activity from producing unnecessary changes in the memristive state. Specifically, learning is initiated only when the effective reinforcement exceeds a predefined learning boundary. The corresponding memristive weight update is expressed as;

$$\Delta \mathrm{W}_{ij}[\mathrm{k}] = \begin{cases} \mathbf{0}, & \mathbf{F_{ij}}[\boldsymbol{k}] < \mathbf{F_{learn}} \\ \boldsymbol{\eta}_{\mathbf{eff,ij}}[\boldsymbol{k}]\left(\mathbf{F_{ij}}[\boldsymbol{k}] - \mathbf{F_{learn}}\right)^{p}, & \mathbf{F_{ij}}[\boldsymbol{k}] \geq \mathbf{F_{learn}} \end{cases} \tag{14}$$

where:

- $\Delta \boldsymbol{W}_{ij}[\mathbf{k}]$ is the positive memristive weight increment associated with sensor pair i,j.
- $\mathbf{F_{ij}}[\boldsymbol{k}]$ represents the effective reinforcement generated by the correlated sensor-derived pulses.
- $\mathbf{F_{learn}}$ is the minimum reinforcement threshold required to initiate learning.
- $\boldsymbol{p}$ is the nonlinear learning exponent governing how rapidly the memristive weight increases beyond the learning threshold.

When the effective reinforcement remains below $\mathbf{F_{learn}}$, no positive learning update occurs, and the incoming sensor activity does not strengthen the corresponding association. Once the reinforcement reaches or exceeds the threshold, the memristive weight increases according to the nonlinear learning function. For $p>1$, reinforcement only slightly above the threshold produces relatively small updates, whereas stronger reinforcement leads to increasingly larger weight changes before any imposed saturation limit. This provides a mechanism for distinguishing weak coincidental sensor activity from more strongly reinforced associations.

## 8. Nonlinear, state-dependent forgetting

In addition to strengthening relevant sensor associations, the proposed framework investigates the incorporation of forgetting as an intrinsic component of the learning process. This is important because a continuously operating system encounters changing environmental conditions, and previously learned associations may not remain relevant indefinitely. Without a forgetting mechanism, repeated learning updates may cause the memory state to become dominated by past observations, potentially limiting its ability to adapt to new conditions. Accordingly, the framework considers a nonlinear forgetting model in which the rate of memory decay depends on the current strength of the stored association. The forgetting dynamics are expressed as;

$$\frac{dW}{dt} = -\lambda W^q \quad (15)$$

where **W** represents the normalized memristive memory state or effective synaptic weight, λ is the forgetting-rate coefficient, and q is the nonlinear forgetting exponent. The negative sign indicates that the stored state decreases in the absence of reinforcement. The exponent q governs the state dependence of the forgetting process. For q=1, the model produces exponential decay, while q>1 introduces a nonlinear decay in which larger memory states initially exhibit a greater absolute rate of change. In contrast, q<1 produces a different nonlinear relaxation behavior, with relatively strong decay of weak states as they approach zero. Importantly, the effective persistence of a learned association depends on both the forgetting parameters and the memory state established through previous reinforcement. The model therefore allows different sensor associations to exhibit different relaxation characteristics depending on their learning histories and the selected forgetting configuration.

In the present simulation study, for the scenario-wise evaluation, the memory state is initialized at each scene boundary. This permits the learning and forgetting behavior associated with individual scenarios to be investigated without carrying previously accumulated memory into the next scene. Furthermore, after removing the input stimulation, the evolution of the stored state is evaluated to examine the independent influence of nonlinear forgetting. The forgetting behavior considered here is a compact analytical representation of the memory dynamics. Its direct physical realization would depend on the retention, relaxation, and switching characteristics of the selected memristive technology.

## 9. Generation of neuronal firing patterns and actuator-control representations

Following reliability-weighted pulse encoding and memristive associative learning, the resulting sensor-dependent memory states contribute to the effective drive of the I&F neuron. The neuron accumulates the influence of the incoming pulses and learned associations, and a binary HIGH/LOW output is generated according to whether the continuous ALC drive exceeds a prescribed firing threshold. The output is represented as;

$$y[k] = \begin{cases} 1, U_{ALC}[k] \geq V_{th} \\ 0, U_{ALC}[k] < V_{th} \end{cases} \quad (16)$$

Where $U_{ALC}[k]$ represents the normalized continuous ALC drive at sample index k, $V_{th}$ is the selected firing threshold, and $y[k]$ is the resulting binary output. The continuous neuronal response is subsequently summarized into a scenario-level analogue output. In the reported simulations, this output is obtained by averaging the continuous ALC drive over each scene, with the contribution of each observation weighted by its associated duration. The resulting normalized analogue value is converted into a numerical 8-bit code.

For each driving scenario, the encoded output is compared with a predefined synthetic reference code representing the desired actuator response. The reference codes are assigned according to the operating condition and the corresponding expected action, while their eventual translation into physical actuator commands would require a dedicated decoding mechanism. This establishes a connection between the underlying encoding and learning parameters and the resulting output representation. Changes in activity-vector size, reliability weighting, pulse amplitude, learning strength, forgetting behavior, or firing threshold may alter the continuous ALC drive and, consequently, its corresponding binary response.

## 10. Evaluation of output-code agreement and information reduction

To investigate the influence of adaptive processing, the simulated output responses are compared with predefined synthetic ground-truth codes across the ten driving scenarios. The primary comparison is based on bit-wise agreement, which measures the fraction of corresponding bits that match between the simulated 8-bit code and its reference code. For a scenario $\mathbf{j}$, bit-wise agreement can be expressed as;

$$\mathbf{B_j} = \frac{1}{8}\sum_{b=1}^{8} \mathbf{1}\left(\hat{\mathbf{z}}_{j,b} = \mathbf{z}^{*}_{j,b}\right) \times 100\% \quad \textbf{(17)}$$

Where $\mathbf{B_j}$ represents the percentage of matching bits for scenario j, $\hat{\mathbf{z}}_{j,b}$ denotes the b-th bit of the simulated output code, and $\mathbf{z}^{*}_{j,b}$ represents the corresponding bit of the predefined reference code.

A higher bit-wise agreement indicates that the generated numerical output more closely reproduces the expected binary representation. Since these binary codes are intended to be interpreted by a pattern decoder, agreement provides an intermediate measure of the correspondence between the ALC output and the desired control representation. However, the eight bits represent a numerical code rather than eight independent temporal firing intervals. Consequently, bit-wise agreement is sensitive to the selected coding scheme and does not directly establish the correctness of the corresponding actuator command. In addition to bit-wise agreement, the mean absolute error (MAE) between simulated and reference analogue outputs is considered to quantify their numerical differences. These two measures provide complementary information, as a configuration producing higher bit-wise agreement does not necessarily produce a lower analogue output error. The effects of individual model parameters are investigated by varying the activity-vector size, near-maximum threshold, number of event-density levels, nonlinear PAM exponent, reliability weighting, memristive learning coefficient, and forgetting parameters relative to a baseline configuration. Finally, the potential advantage of scenario-adaptive activity-vector sizing is evaluated by comparing the retained information against fixed-vector configurations. Assuming 32-bit representation for each activity point, the retained payload per synchronized observation is determined from the activity-vector size and the three sensor modalities. This comparison allows the study to investigate whether reducing the retained sensory information necessarily leads to deterioration in output-code agreement.

# 3. Results and Discussions

To evaluate these capabilities under representative driving conditions, the publicly available nuScenes dataset [42] is used to construct a scenario-specific ground-truth reference, as summarized in Table 1. The selected dataset observations are organized into ten predefined scenarios, each spanning a continuous range of synchronized sample indices. For instance, Scenario 1 extends from sample index 1 to 39, representing an operating condition for which an appropriate driving response can be established from the observed environmental context. Each scenario is subsequently assigned an expected 8-bit ALC firing pattern, representing a predefined temporal sequence of binary neuronal responses that can be interpreted by a spike-pattern detector and mapped to actuator commands, such as braking, steering, or maintaining the current driving state. The simulated ALC outputs are then compared with these reference patterns to quantify decision accuracy across different scenarios and parameter configurations.

**Table 1. Ground-truth outcomes and expected ALC output patterns for representative scenarios:** A summary of the expected true outcomes for each scenario of NuScenes, including the corresponding temporal HIGH/LOW output pattern of the ALC that will ultimately drive the actuator. serving as the reference for evaluating all subsequent results.

| Sample range | Scenario / physical condition | Expected ALC state | Ground-truth output pattern within decision window | Actuator interpretation | Expected decision |
|---|---|---|---|---|---|
| 1–39 | Scene 1<br>Road approach; parked truck left, construction barriers right, van ahead -> Intersection/crosswalk approach and leftward route behind van -> Following van through cone/barrier corridor | LOW request throughout; monitoring may remain active → Sustained HIGH request | **00111011** | No active slowing/yield/hold request. Reduce accelerator and follow intended left turn, | Continue cautiously -> Slow down; follow leftward route -> Slow down; maintain following gap |
| 40–79 | Scene 2<br>Intersection approach; opposing traffic and pedestrians near sidewalks -> Dark vehicle turns across lane ahead and then travels ahead -> Lane ahead opens; parked cars and opposing vehicles remain | LOW request throughout; monitoring may remain active → LOW -> HIGH -> LOW request | **00011000** | Active slowing/hold request; continue monitoring. Monitor intersection and surrounding traffic. Deaccelerate/ brake if required to yield, release the break as path clears | Continue cautiously -> Yield / slow down, then follow -> Continue cautiously |
| 80–120 | Scene 3<br>Waiting at intersection while vehicles and large vehicle cross -> Multiple pedestrians including stroller cross directly in front -> Last pedestrians leave foreground crosswalk; intersection still under observation | Sustained HIGH request | **01101100** | Maintain stopped state while crosswalk is occupied. | Hold stopped / yield -> Hold stopped; yield to pedestrians -> Hold pending right of way |
| 121–161 | Scene 4<br>Urban road; parked cars and opposing vehicles -> Crosswalk approach; pedestrian near/crossing left side and bus ahead right -> Passing bus to right and work vehicles/cones to left | LOW request throughout; monitoring may remain active → Sustained HIGH request | **00101010** | No active slowing/yield/hold request; continue monitoring. Maintain lane, monitor opposing traffic, Temporary slowing/yield request in middle intervals, release afterwards. Reduce accelerator approaching crossing | Continue cautiously/Slow down and Continue at reduced speed |
| 162–202 | Scene 5<br>Approach busy intersection beside construction equipment -> Bus and repeated vehicles cross ahead; ego appears waiting -> Continued cross traffic at intersection | LOW -> HIGH request → Sustained HIGH request → Sustained HIGH request | **00110011** | Slowing request begins in second half and remains active at window end. Reduce accelerator approaching intersection. | Slow down to stop |
| 203–242 | Scene 6<br>Daytime road and junction approach; roadside pedestrians/signs -> Proceeding through junctions with apparently open lane -> School-zone surroundings and blue car drawing alongside on right | LOW request throughout; monitoring may remain active | **00101010** | No active slowing/hold request. This does not command acceleration. Maintain lane and monitor junction. Smooth cruising | Continue cautiously monitor adjacent vehicle |
| 243–283 | Scene 7<br>Low-speed parking-area approach; pedestrian close ahead/left -> Parking aisle with parked cars on both sides -> Exit/turn near parked vehicles and roadway | HIGH -> LOW request → Sustained HIGH request → Sustained HIGH request | **01100011** | Slowing request active initially, released in second half as the proposed conflict clears. Limit speed and monitor nearby pedestrian. Maintain low speed and lateral clearance. Follow route at low speed and check right of way | Slow down before entering road |
| 284–324 | Scene 8<br>Night road following car; glare and traffic in other lanes -> Continued night driving past roadside work/cones -> Night road and bus-stop/roadside features; lead vehicle ahead | LOW request throughout; monitoring may remain active | **10001100** | No active slowing/hold request, continue monitoring. Maintain following distance and monitor lead car. Smooth cruising. | Continue cautiously |
| 325–364 | Scene 9<br>Wet night junction; white lead car and bright reflections -> Pedestrian crosses in foreground as lead vehicle moves away -> Following car on night road; pedestrians mostly roadside | LOW request throughout; monitoring may remain active | **10001011** | Reduce accelerator, brake/hold while path occupied. | Slow down / follow cautiously |
| 365–404 | Scene 10<br>Night intersection with repeated crossing/turning cars; ego appears waiting -> More cross traffic and small two-wheel vehicles across intersection -> Continued vehicle crossings; nighttime glare | Sustained HIGH request | **01001100** | Request remains active throughout the window, action type is specified by expected decision, not bit count. Maintain hold for conflicting vehicles. Continue hold pending until confirmed right of way | Hold stopped / yield -> Hold stopped; yield to pedestrians -> Hold pending right of way |

The first objective of this analysis is to investigate the severity of the operating conditions across these ten selected driving scenarios (Scenes 1–10), which ultimately serves as the basis for adapting the other parameter functions of the proposed framework. The scenario complexity is computed using the model described in previous section. Figure 3 illustrates the variations in pairwise sensor-association strengths, the four contributing parameters, and the resulting complexity across all 404 samples. Based on the computed complexity, the operating conditions are categorized into three levels: low, moderate, and high, as highlighted in Figure 3. It can be clearly observed that the complexity varies considerably, not only between different scenarios but also within individual scenarios, depending on the changing sensory conditions. For instance, relatively high complexity is observed during certain intervals of Scenes 1, 6, and 7, whereas other intervals exhibit lower complexity. Importantly, **each sample index retains its own complexity level, determined independently using Eq. (3), rather than assigning a fixed complexity to an entire scenario.** This sample-wise assessment is particularly important for the proposed adaptive framework, as it allows the subsequent encoding, reliability, and learning parameters to be adjusted according to the continuously changing severity of the encountered situation. The values of the parameters that are used in the simulations are mentioned as follows;

| **Parameters** | **Resolved Values used in Simulations** |
|---|---|
| Complexity weights $w_1, w_2, w_3$ and $w_4$ | 0.20, 0.20, 0.40, 0.20 |
| N and density count | 8/16/64; 1/2/3 levels |
| X | Continuous function of scene-specific mean/std and frame complexity |
| PAM | Power law, $\gamma$=1.5 in all modes; $\alpha_{temporal}$=3, $\beta_{temporal}$=2 |
| Reliability | Raw-quality gate × agreement/temporal consistency |
| Learning | $\eta_0$=0.05; $F_{learn}$=0.10; p=1.5; $W_{max}$=1; $\tau$=0.05 s |
| Forgetting | $\lambda_0$=0.02; q=0.50; $\lambda_{effective}=\lambda_0*(1.5-C_{sc})$ |
| Output | Recall gain 0.50; firing 0.35 |

After establishing the scenario complexity, the activity-vector size is adaptively adjusted according to the complexity level of each sample index. As illustrated in Figure 4, the activity-vector size increases from (N=8) for a low-complexity sample (C=0.184) to (N=16) for a moderate-complexity sample (C=0.411), and further to (N=64) for a high-complexity sample (C=1.000). This adaptation is motivated by the fact that more severe or complex situations generally require a greater amount of sensory information to support reliable decision-making, while relatively simple situations can be effectively handled using a limited amount of information. To compute the event density, the near-maximum activity threshold (X) is also made adaptive to the scenario complexity. This adjustment is motivated by the fact that more complex or severe situations may require information from a greater number of activity points to support reliable decision-making, whereas relatively simple situations can be effectively represented using fewer highly active points. Figure 5 illustrates the variation in mean scenario complexity across the ten selected scenarios, along with the corresponding adaptation of (X) for Camera, LiDAR, and Radar. It can be observed that the selected thresholds generally decrease as complexity increases, although their values differ across sensor modalities according to their respective activity statistics. Thus, the adaptive threshold enables scenario-dependent control over the amount and diversity of sensor activity retained for event-density computation, providing an additional level of information optimization beyond the adaptive activity-vector size. The number of event-density levels is also made adaptive, with higher-complexity samples assigned a larger number of levels and lower-complexity samples represented using fewer levels, as illustrated in Figures 6 and 7. This allows complex situations to retain richer information about the spread of activity within the vector, whereas simple situations can still be described compactly without unnecessary processing overhead. Figure 6 shows that, across all 404 observations, low-complexity samples are predominantly represented using a single event-density level, moderate-complexity samples more often require two levels, and high-complexity samples are generally associated with three levels. The scene-wise distribution further confirms that the selected number of levels varies according to the complexity characteristics of each scene. Representative examples are shown in Figure 7, where high-complexity scenes, such as Scenes 1 and 6, are described using three event-density levels, while low-complexity scenes, such as Scenes 3 and 8, are adequately represented using only one level. The resulting continuous composite event density, (E), which varies at every sample index as a function of the corresponding complexity level (C) and adaptive threshold (X), is subsequently used to generate PAM signals according to Eq. (10). The corresponding nonlinear power mapping and the resulting PAM amplitudes for representative scenes are illustrated in Figure 8, showing how different event-density profiles lead to different pulse-amplitude behaviors across the sample indices. As a representative example, Figure 9 illustrates the generated nonlinear PAM pulse trains for Camera, LiDAR, and Radar in Scene 7, where each synchronized sample index is represented by a single pulse with an amplitude determined by its corresponding composite event density and temporal variations. In particular, Camera and LiDAR exhibit relatively higher pulse amplitudes during certain intervals, whereas Radar produces comparatively weaker responses throughout the scenario. These variations demonstrate how the proposed nonlinear encoding converts continuously changing multimodal sensor activity into compact, amplitude-modulated pulse representations while preserving differences in the relative strength and temporal behavior of the sensor-derived information. The resulting pulse trains subsequently serve as inputs to the reliability-aware processing stage before being applied to the associative learning circuit. As illustrated in Figure 10, that the estimated reliability varies across the three sensor modalities, directly influencing the amplitude of the pulses retained after reliability weighting. In particular, LiDAR exhibits consistently high confidence, resulting in relatively minor attenuation of its generated pulses, whereas Radar shows considerably lower confidence, leading to substantial attenuation of its pulse amplitudes. Camera exhibits comparatively high but varying confidence, resulting in moderate, sample-dependent attenuation. This adaptive adjustment ensures that sensor activities with higher reliability retain a stronger influence on the subsequent processing stages, while less reliable observations contribute with reduced strength. Thus, reliability-weighted pulse encoding provides an additional level of control over the information supplied to the associative learning circuit, limiting the influence of uncertain sensor observations without necessarily discarding them entirely. The key objective of reliability-aware processing is to control the learning of the memristive element within the neuron circuit, as its conductance state ultimately influences the neuronal firing pattern. Figure 11 illustrates the estimated pairwise confidence for Camera–LiDAR, Camera–Radar, and LiDAR–Radar, along with their corresponding contributions to memristive weight reinforcement $\Delta W$. It can be observed that the Camera–LiDAR pair exhibits relatively high confidence, resulting in stronger reinforcement, whereas the Camera–Radar and LiDAR–Radar pairs have lower confidence and consequently contribute less to the learning process. In particular, the LiDAR–Radar pair exhibits comparatively low reinforcement, despite occasional increases in its unweighted learning contribution. This allows reliable relationships to be strengthened while limiting the influence of uncertain or potentially corrupted observations on subsequent neuronal firing behavior. Another important aspect investigated in this study is the incorporation of nonlinear forgetting into the memristive associative learning process, as described in Eq. (15). Figure 12 illustrates the combined influence of learning and forgetting on the evolution of the memristive state for the three sensor pairs: Camera–LiDAR, Camera–Radar, and LiDAR–Radar. The left panels show the evolution of the normalized memristive conductance across all 404 sample indices, while the middle panels illustrate the corresponding learning dynamics resulting from the simultaneous reinforcement and forgetting of the stored associations. In this analysis, the memristive state is assumed to reset at each scene boundary, allowing the learning behavior of individual scenarios to be evaluated independently. To further investigate the influence of nonlinear forgetting, the right panels illustrate the decay of the learned associations after removing the input pulses, considering the different complexity levels and memory states established in each scene. It can be observed that the forgetting behavior varies considerably across the scenarios, with weaker associations generally decaying faster, while more strongly reinforced associations persist for longer durations.

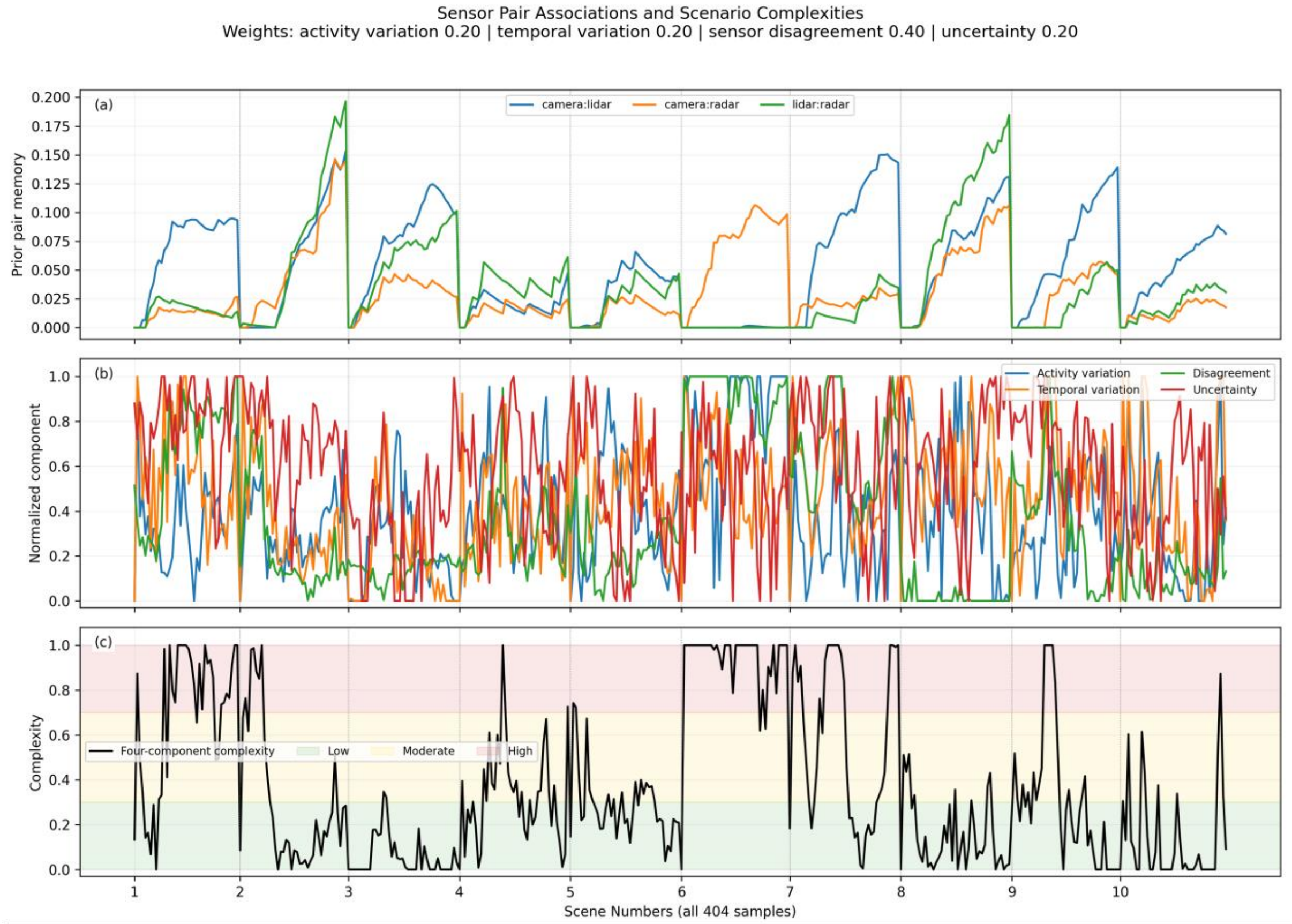

**Figure 3: Scenario complexity estimation across ten driving scenarios. Evolution of (a) pairwise sensor-association strengths for Camera–LiDAR, Camera–Radar, and LiDAR–Radar; (b) four normalized complexity indicators, including activity variation, temporal variation, cross-sensor disagreement, and uncertainty; and (c) the resulting sample-wise scenario complexity across 404 synchronized observations, classified into low, moderate, and high complexity regions.**

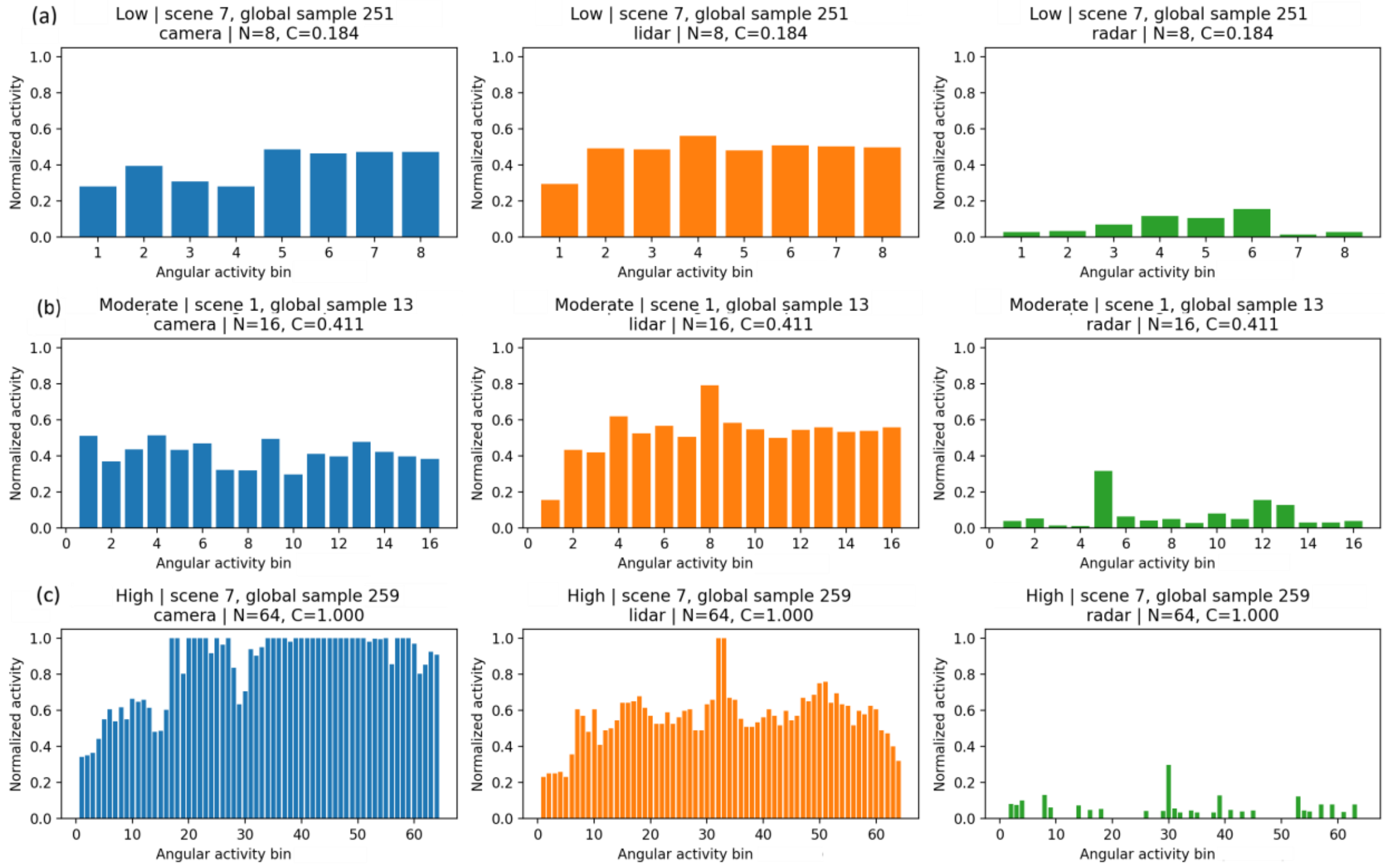

**Figure 4. Scenario-adaptive activity-vector sizing for multimodal sensor inputs. Representative Camera, LiDAR, and Radar activity distributions under (a) low ((C=0.184), (N=8)), (b) moderate ((C=0.411), (N=16)), and (c) high ((C=1.000), (N=64)) complexity conditions, demonstrating the adaptation of retained activity-vector resolution according to scenario severity.**

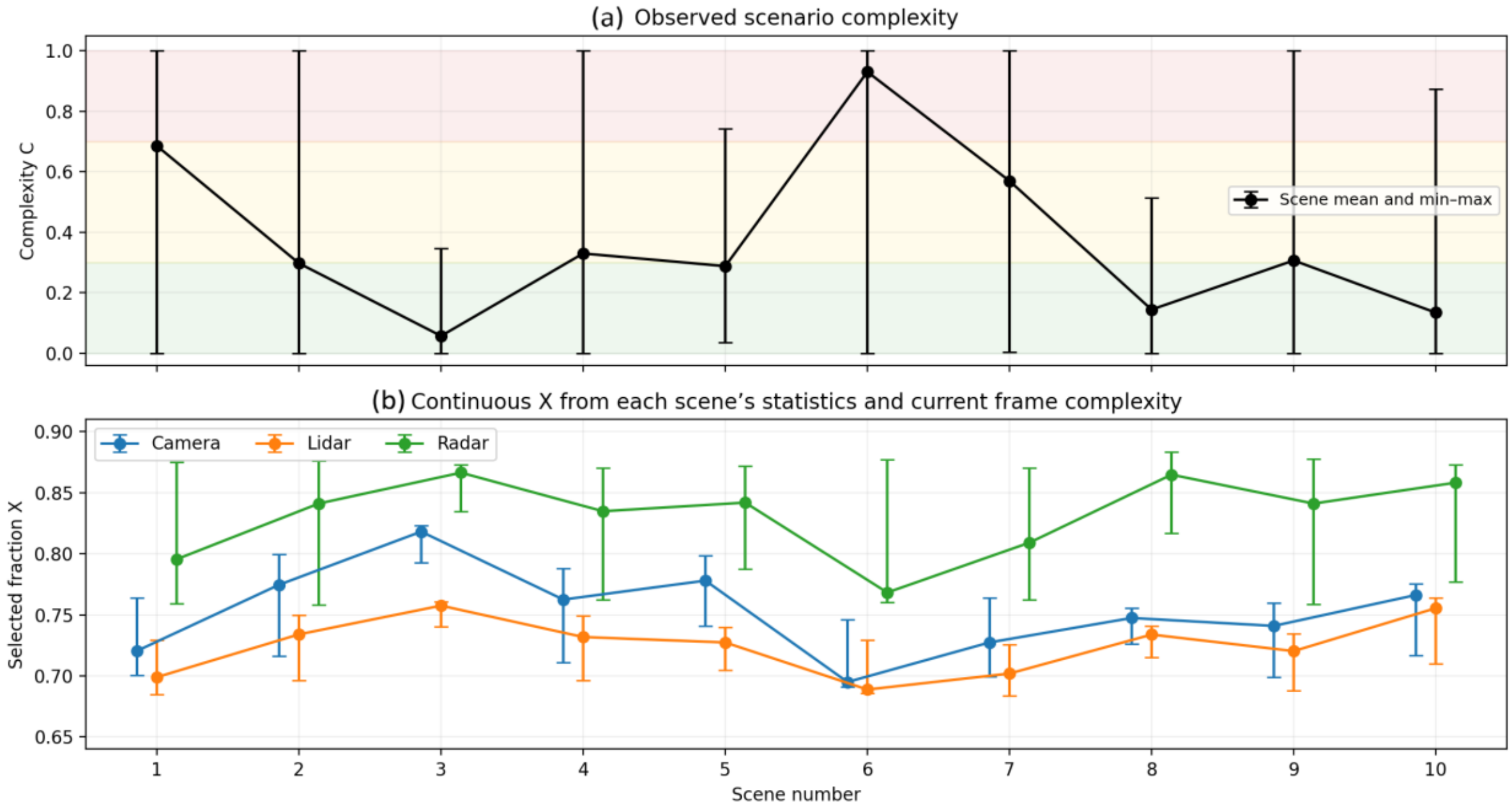

**Figure 5. Adaptive near-maximum activity threshold selection. (a) Mean scenario complexity with minimum–maximum variations across the ten driving scenarios and (b) corresponding near-maximum activity thresholds ((X)) for Camera, LiDAR, and Radar, illustrating the inverse dependence of the selected thresholds on scenario complexity and their variation with sensor-specific activity statistics.**

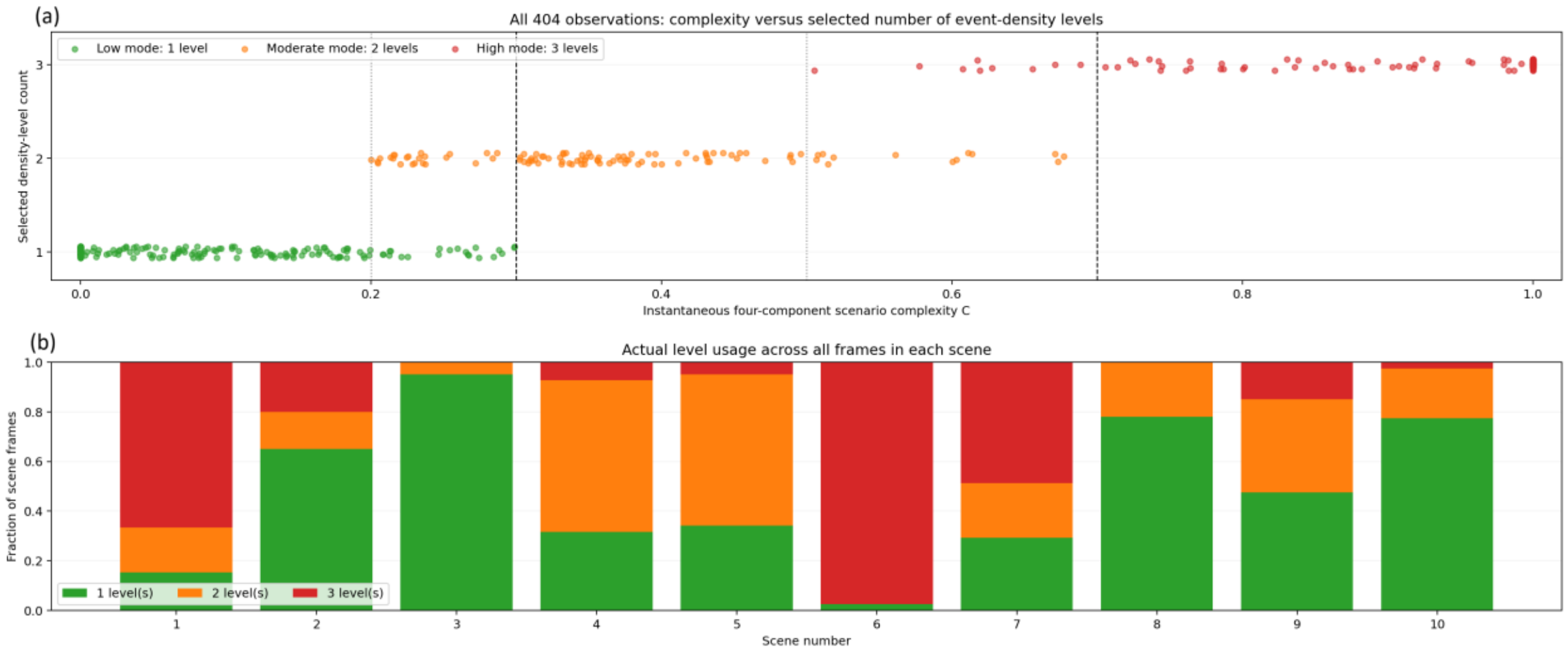

**Figure 6. Scenario-dependent selection of event-density representation levels. (a) Distribution of the selected number of event-density levels across all 404 observations as a function of instantaneous scenario complexity and (b) scene-wise proportions of single-, two-, and three-level representations, demonstrating adaptive information resolution according to operating conditions.**

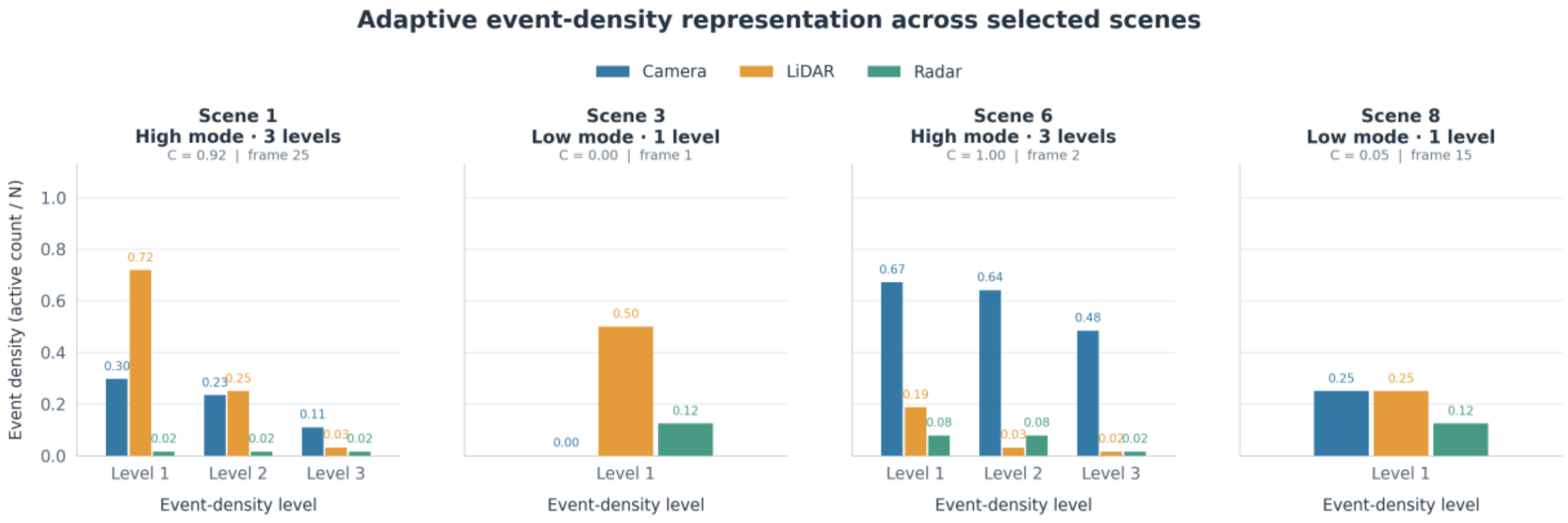

**Figure 7. Multilevel event-density representations under different scenario complexities. Representative event-density values for Camera, LiDAR, and Radar in Scenes 1, 3, 6, and 8, illustrating the use of three levels for high-complexity observations and a single level for low-complexity observations.**

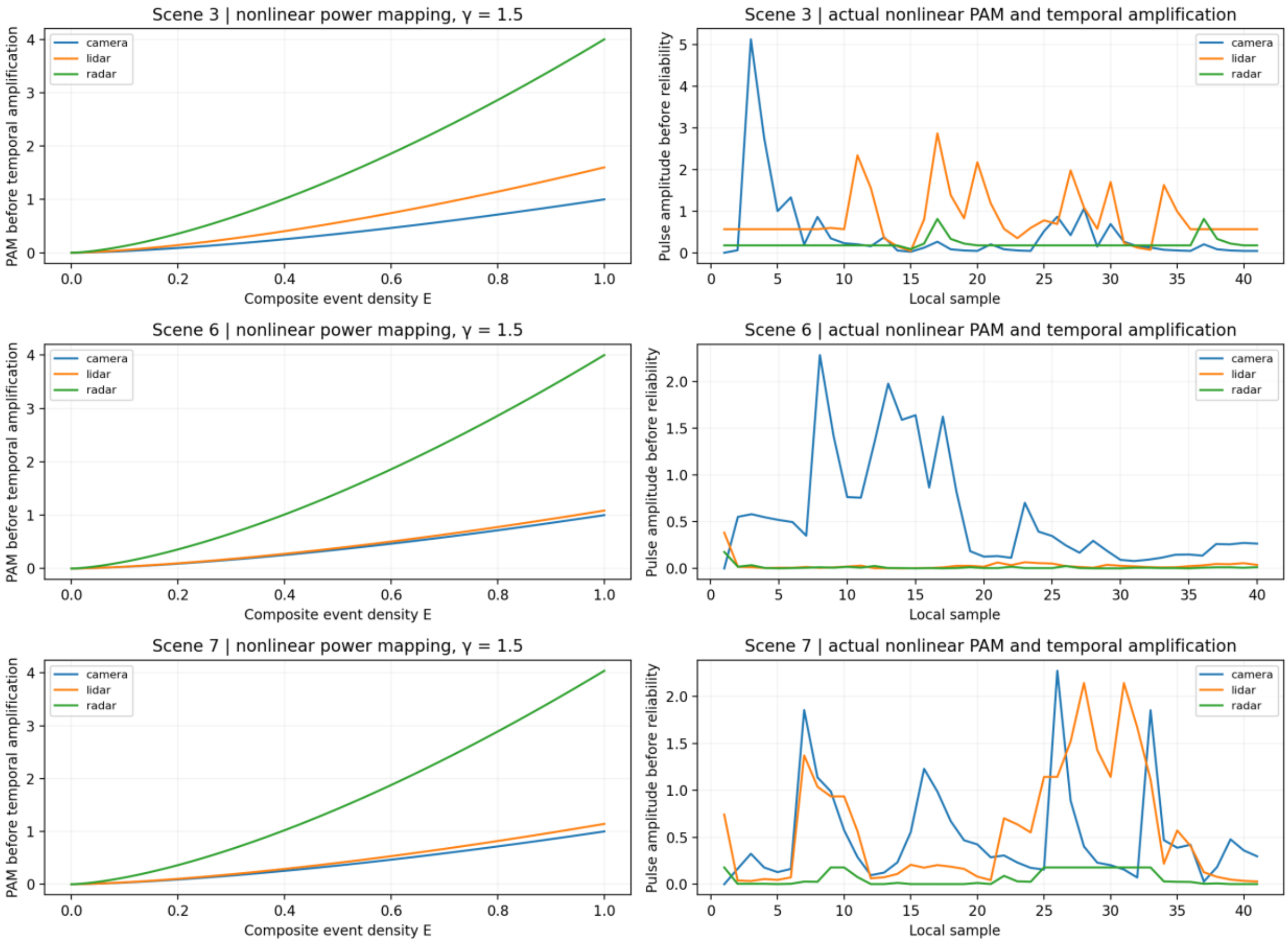

**Figure 8. Nonlinear pulse-amplitude modulation of composite event density. Nonlinear mapping between composite event density and PAM amplitude (left panels), together with the resulting temporally amplified pulse amplitudes (right panels) for Camera, LiDAR, and Radar in representative Scenes 3, 6, and 7, considering a nonlinear exponent of $(\gamma=1.5)$.**

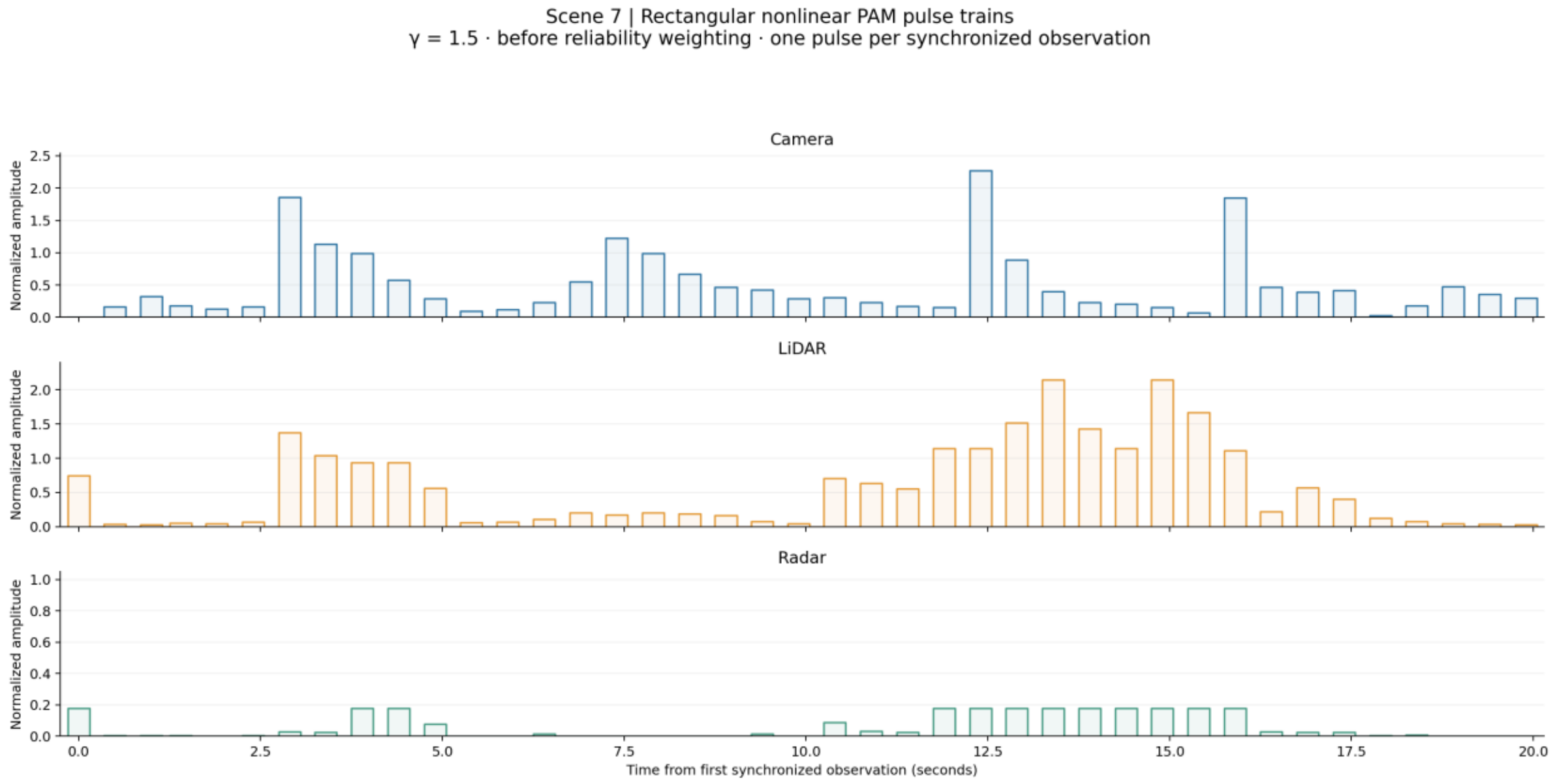

**Figure 9. Nonlinear PAM pulse trains generated from multimodal sensor activity. Representative rectangular PAM pulse trains for Camera, LiDAR, and Radar in Scene 7, showing the sample-dependent variation in normalized pulse amplitudes resulting from composite event density and its temporal changes, before reliability weighting.**

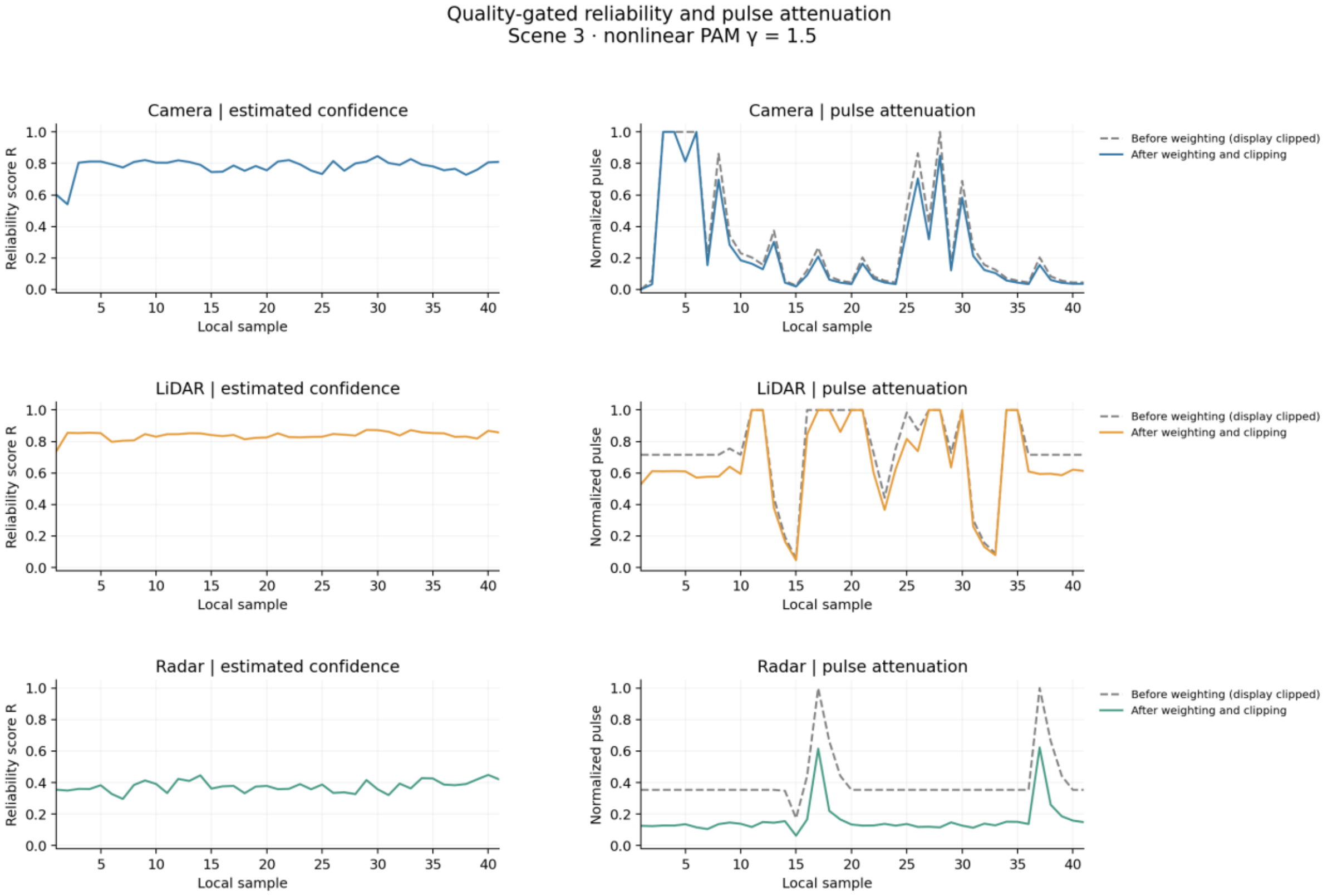

**Figure 10. Reliability-aware attenuation of sensor-encoded PAM pulses. Estimated sensor confidence (left panels) and corresponding PAM amplitudes before and after reliability weighting (right panels) for Camera, LiDAR, and Radar in Scene 3, illustrating confidence-dependent attenuation of unreliable sensor activity.**

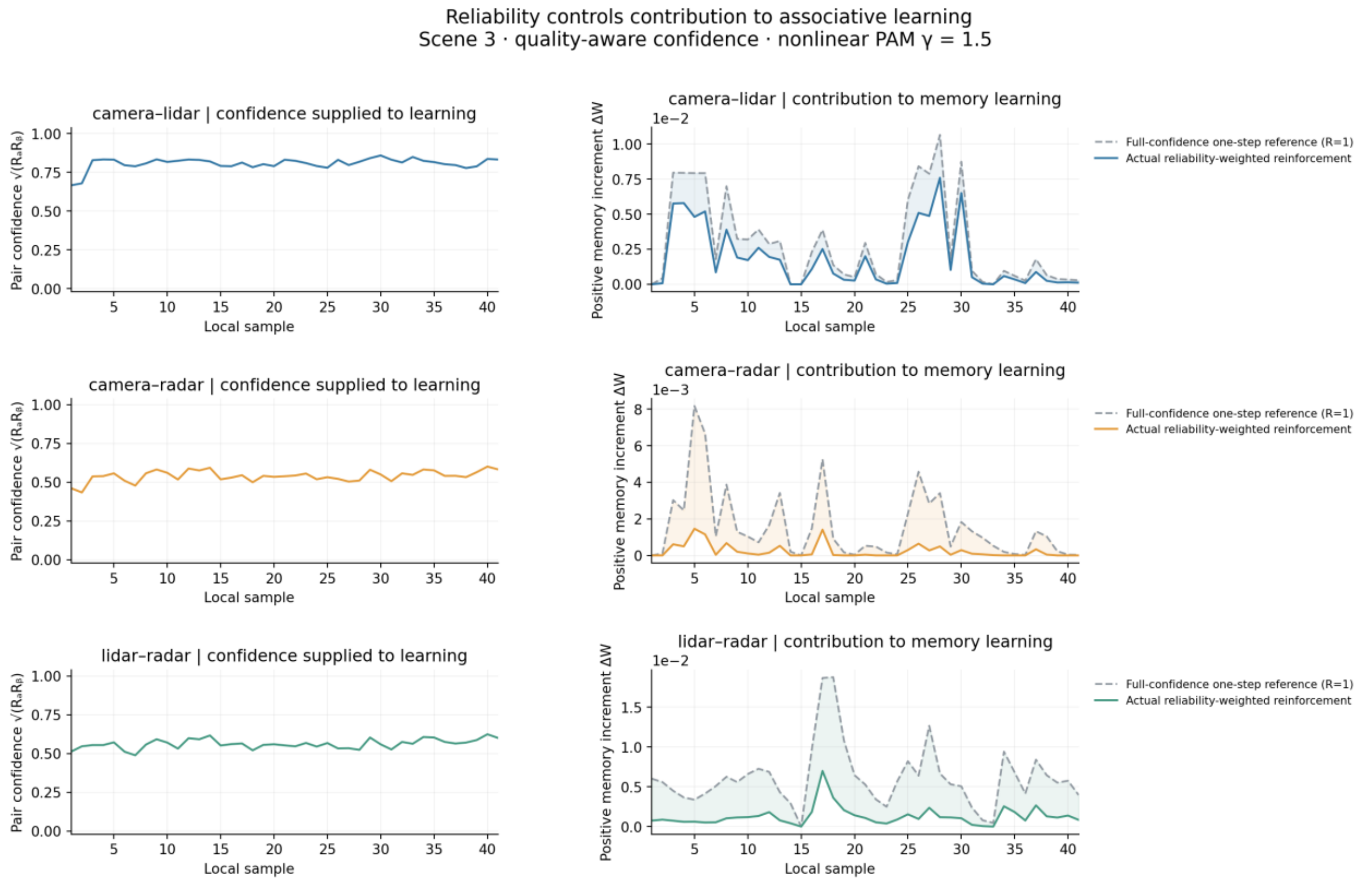

**Figure 11. Influence of sensor-pair reliability on associative learning. Estimated pairwise confidence for Camera–LiDAR, Camera–Radar, and LiDAR–Radar (left panels), and the corresponding positive memristive weight updates compared with full-confidence reinforcement (right panels), demonstrating reliability-dependent regulation of synaptic learning.**

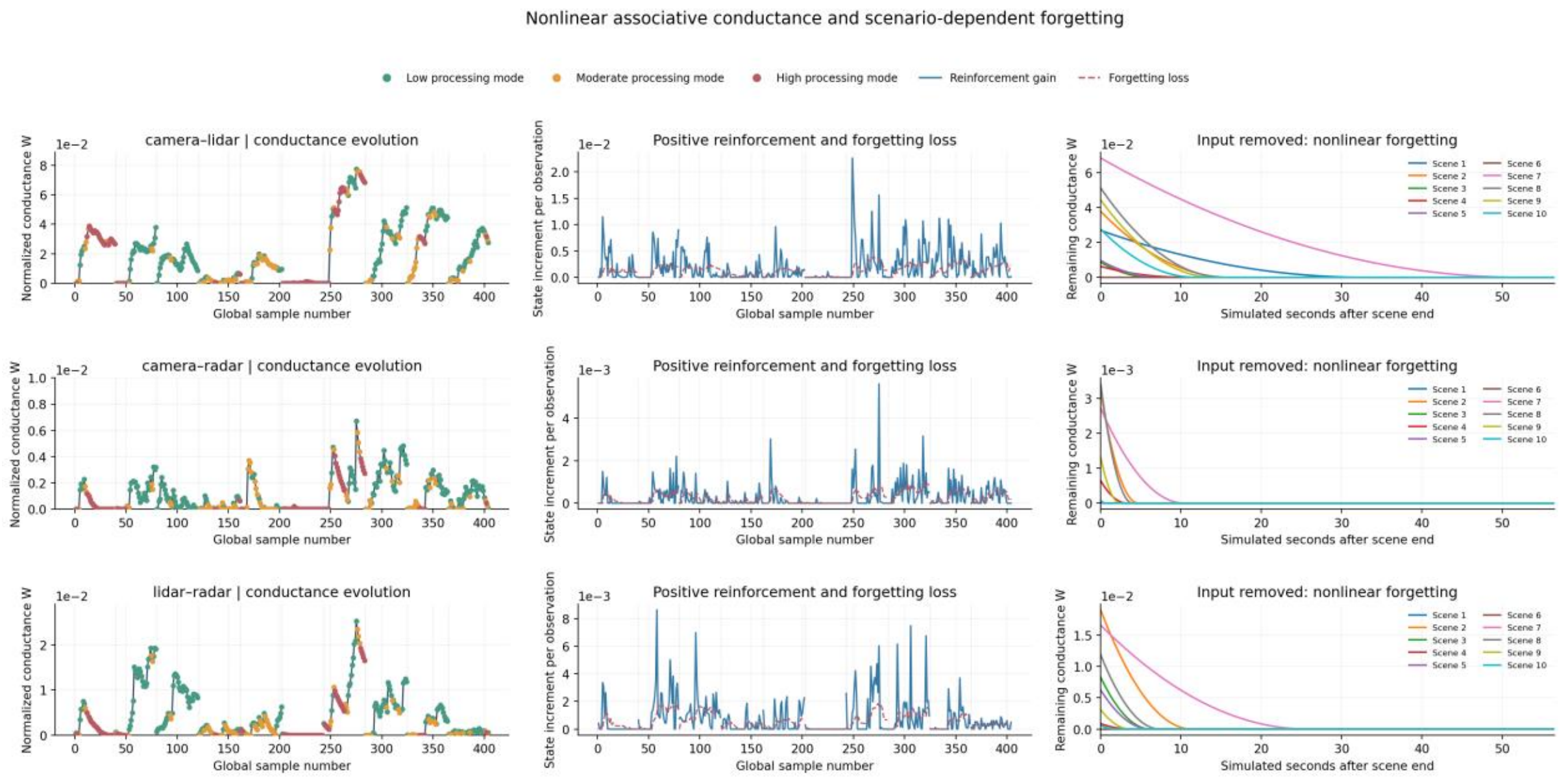

**Figure 12. Nonlinear memristive associative learning and scenario-dependent forgetting. Evolution of normalized memristive states for the three sensor pairs across 404 observations (left panels), corresponding positive reinforcement and forgetting contributions (middle panels), and nonlinear decay of the learned associations following removal of input stimulation for the ten independently evaluated scenarios (right panels).**

The influence of the neuronal firing threshold on the output response of the ALC is further investigated by considering two fixed threshold values of 0.40 and 0.35, as illustrated in Figures 13(a) and 13(b), respectively. The upper panels show the variation in the continuous ALC drive across all 404 sample indices, while the lower panels represent the corresponding binary HIGH/LOW firing patterns generated when the drive crosses the selected threshold. It can be observed that reducing the threshold from 0.40 to 0.35 increases the sensitivity of the neuron, allowing relatively weaker input activity to trigger additional firing events. Finally, the simulations are conducted to systematically evaluate the influence of various processing and learning parameters on the ALC output patterns, using a fixed baseline configuration for comparison. The objective is to investigate whether allowing these parameters to adapt according to the changing complexity of the operating environment can improve decision accuracy, particularly in terms of agreement with the predefined ground-truth firing patterns. Figure 14 illustrates the variations in the binary HIGH/LOW output responses across all 404 sample indices when individual parameters, including activity-vector size, near-maximum activity threshold, event-density representation, nonlinear PAM encoding, reliability level, memristive learning strength, and forgetting behavior, are modified relative to their baseline values. It can be observed that changes in these parameters produce different firing patterns, with certain configurations increasing neuronal activation while others suppress firing events during specific intervals. These variations demonstrate that ***the output of the ALC is strongly influenced by the combined behavior of the encoding, reliability, and synaptic learning stages, rather than by the firing threshold alone. The analysis therefore highlights the potential of scenario-adaptive parameter optimization to achieve better agreement with the expected actuator-response patterns, while avoiding unnecessary information processing and neuronal activations under relatively simple operating conditions.***

Finally, the simulated ALC responses are compared with the predefined ground-truth patterns. Figure 15 compares the output patterns obtained using the adaptive parameter configuration with the corresponding ground-truth codes. It can be observed that complete agreement is achieved only for Scene 4, whereas the remaining scenarios exhibit varying degrees of mismatch. Overall, the simulation achieves a mean bit-wise agreement of 61.2%, with only one out of ten scenes producing an exact match with the predefined reference code. The results therefore indicate that, although adapting the encoding, reliability, and memristive learning parameters influences the ALC firing behavior, the current adaptive configuration is not yet sufficiently optimized to reproduce the desired output patterns consistently across different scenarios. This highlights the need for further investigation into the combined optimization of these parameters and the design of an appropriate pattern-decoding mechanism to improve the correspondence between neuronal firing responses and the intended actuator-control decisions.

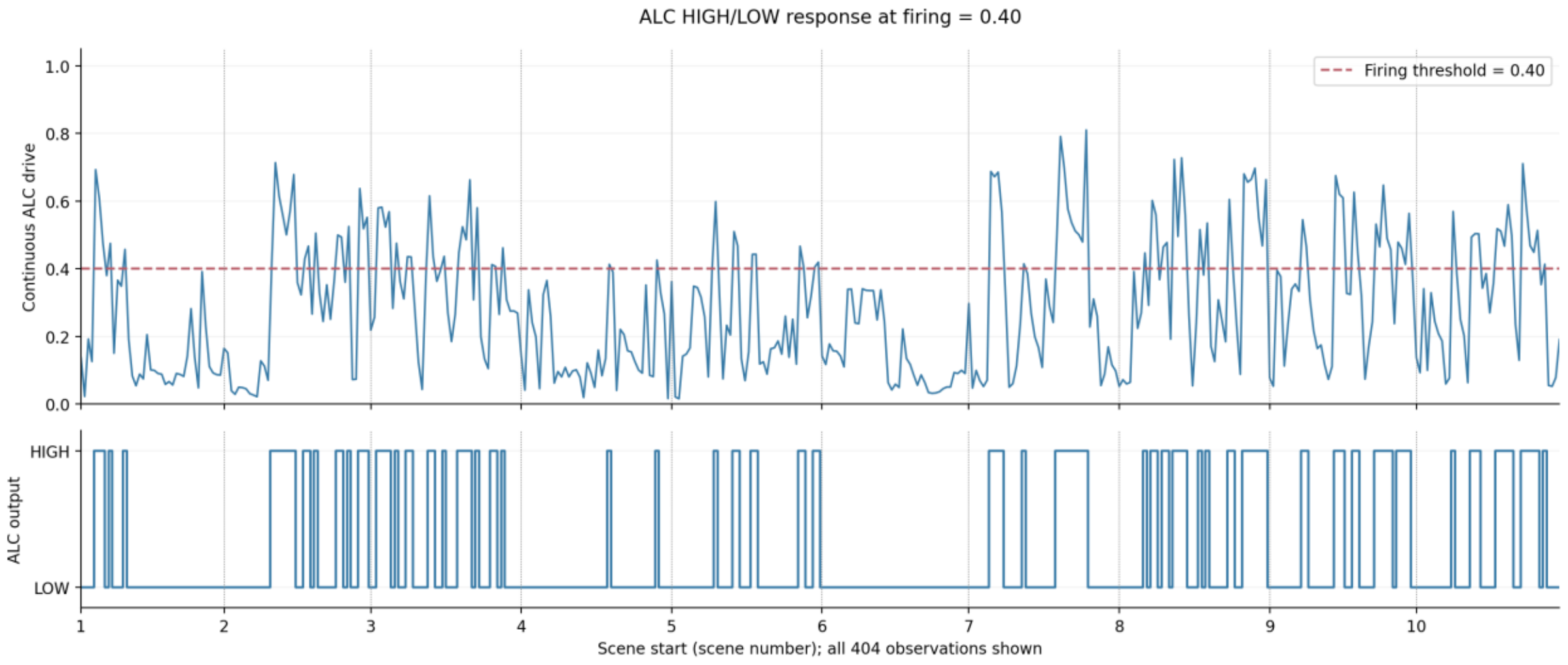

(a)

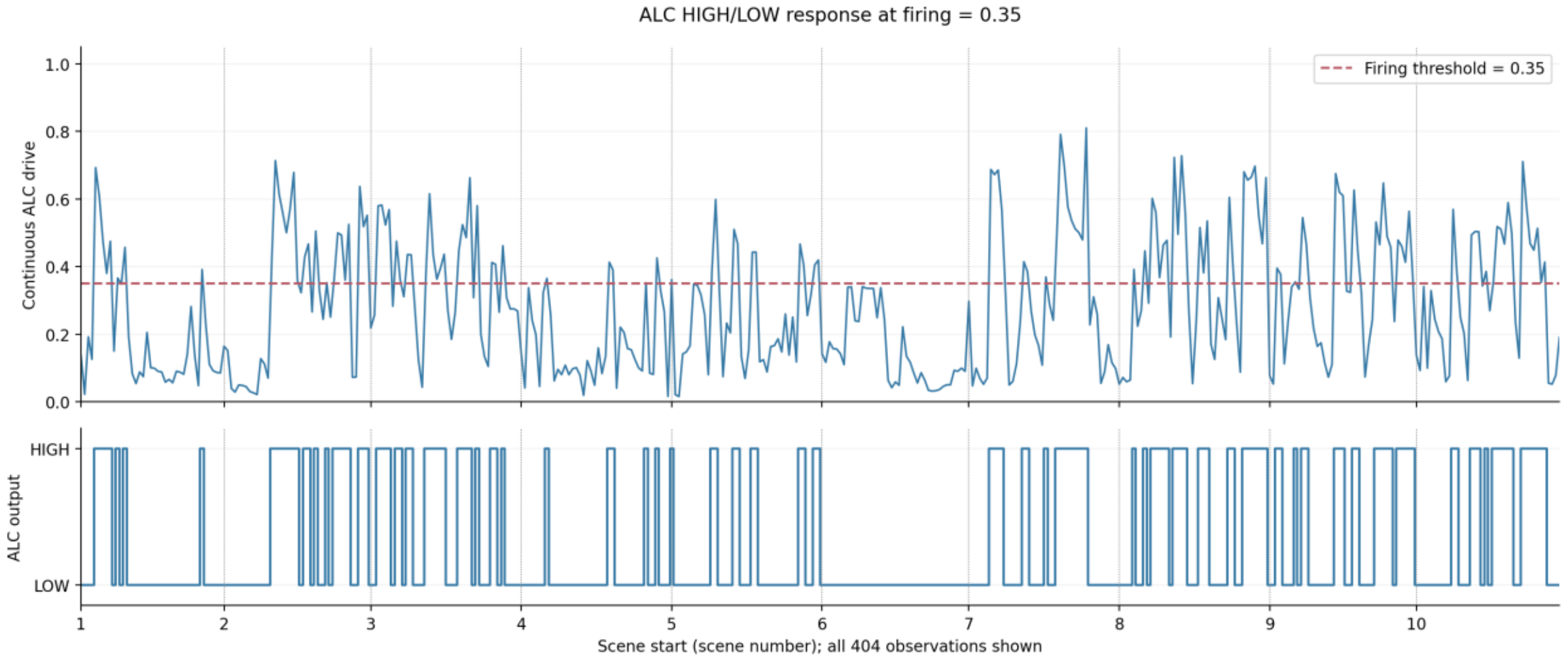

(b)

**Figure 13. Influence of firing threshold on the binary output response of the associative learning circuit. Continuous ALC drive and corresponding HIGH/LOW firing patterns across 404 observations for fixed firing thresholds of (a) 0.40 and (b) 0.35, illustrating the effect of neuronal sensitivity on the generated output sequences.**

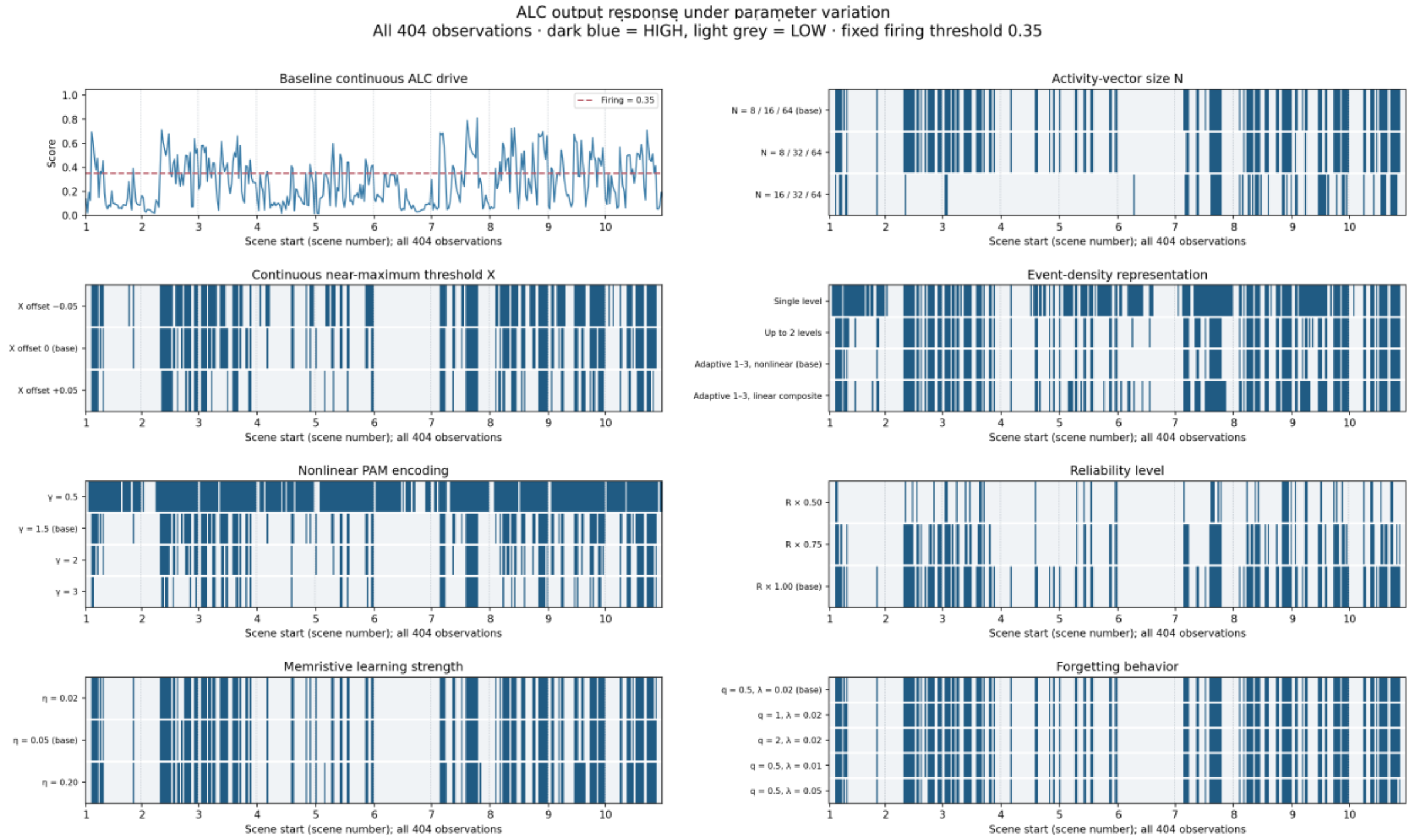

**Figure 14. Sensitivity of ALC firing patterns to encoding and learning parameters. Comparison of binary HIGH/LOW responses across 404 observations under variations in activity-vector size, near-maximum threshold, event-density representation, nonlinear PAM encoding, reliability level, memristive learning strength, and forgetting behavior, relative to their respective baseline configurations**

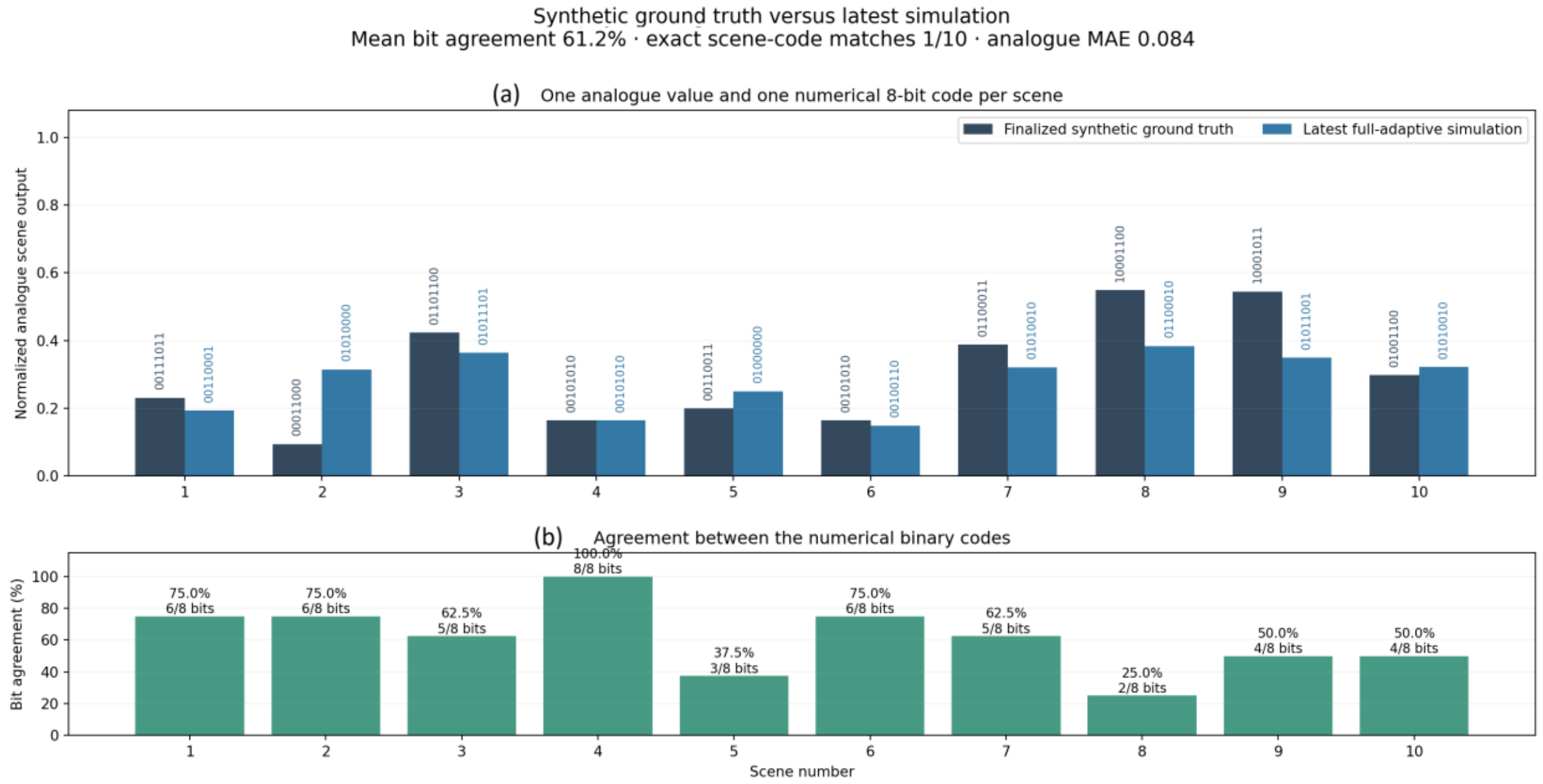

**Figure 15. Comparison of adaptive ALC responses with synthetic ground-truth output codes. (a) Scene-wise normalized analogue outputs and corresponding numerical 8-bit codes for the adaptive simulation and predefined ground truth, and (b) bit-wise agreement across ten driving scenarios, yielding a mean agreement of 61.2% and one exact scene-code match out of ten.**

To further investigate the influence of individual parameters on the decision-making capability of the ALC, Figure 17 presents a quantitative comparison of the generated output patterns with the predefined synthetic ground-truth codes. Bit-wise agreement is particularly important in this analysis because it indicates how closely the ALC output reproduces the binary code assigned to the desired actuator response for each driving scenario. A higher agreement therefore suggests a closer correspondence between the neuronal response and the expected control representation, which is an important requirement for reliable spike-pattern decoding. The unchanged adaptive baseline achieves a mean bit-wise agreement of 61.2%, providing a reference for investigating the influence of different encoding, reliability, and memristive learning parameters. It can be observed that the agreement is particularly sensitive to the input encoding and reliability settings. For instance, increasing the nonlinear PAM exponent from 1.5 to 3 reduces the agreement from 61.2% to 47.5%, whereas modifying the reliability weighting to 0.75 improves it to 65.0%. In comparison, variations in memristive learning strength and forgetting behavior produce relatively smaller changes, with the latter achieving a maximum agreement of 63.8% under the investigated configurations. These results highlight that the ability of the neuron circuit to generate the expected output representation is strongly influenced by how sensory information is compressed, encoded, and allowed to participate in learning. Importantly, since the output patterns are intended to be interpreted by a dedicated pattern decoder for actuator control, improving bit-wise agreement represents a step toward generating more consistent scenario-specific responses. However, bit-wise agreement alone does not establish the correctness of the final actuator decision, as this also depends on the code assignments and decoding mechanism. Overall, the results demonstrate the importance of coordinated parameter optimization in improving the correspondence between hardware-generated neuronal responses and the intended actuator-control patterns.

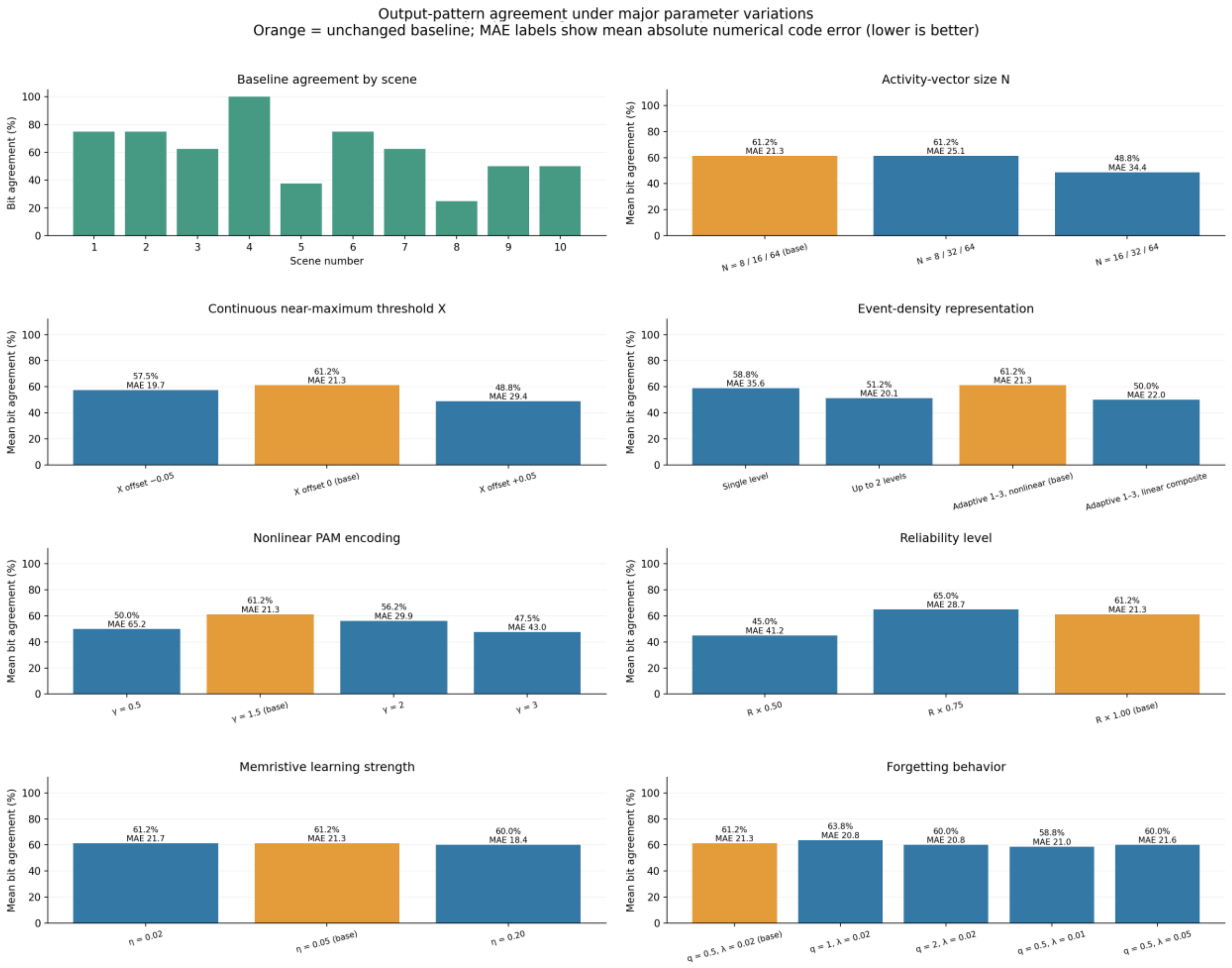

**Figure 16. Sensitivity of ALC output-pattern agreement to processing and learning parameters: Scene-wise bit agreement for the baseline configuration and comparison of mean bit-wise agreement with predefined synthetic ground-truth codes under variations in activity-vector size, near-maximum activity threshold, event-density representation, nonlinear PAM encoding, reliability weighting, memristive learning strength, and forgetting behavior. Orange bars indicate the unchanged adaptive baseline, while blue bars represent alternative parameter configurations. The corresponding mean absolute numerical code errors (MAE) are indicated above the bars, illustrating the sensitivity of the ALC output response to individual parameter variations.**

Figures 17(a) and 17(b) further investigate the trade-off between information reduction and agreement with the predefined synthetic ground-truth codes under different activity-vector sizes. It can be observed that scenario-adaptive activity-vector sizing achieves a 61.5% reduction in the retained sensor-data representation, reducing the payload from 6,144 bits to 2,367 bits per synchronized observation compared with the fixed (N=64) configuration. Interestingly, despite processing substantially less information, the adaptive configuration improves the output-code bit agreement from approximately 55% to 61.2%, while also reducing the mean absolute analogue output error from approximately 0.156 to 0.084. In contrast, the fixed activity-vector configurations with (N=32), (N=16), and (N=8) achieve progressively greater information reduction but do not provide a corresponding improvement in output agreement. These results suggest that retaining more sensory information does not necessarily lead to better agreement with the expected output patterns, and that adapting the amount of information according to scenario complexity can provide a more favorable balance between processing requirements and output-code agreement. Although the overall agreement remains limited, this analysis highlights the potential of scenario-adaptive compression to reduce the sensor-data representation without necessarily compromising the quality of the resulting neuromorphic responses.

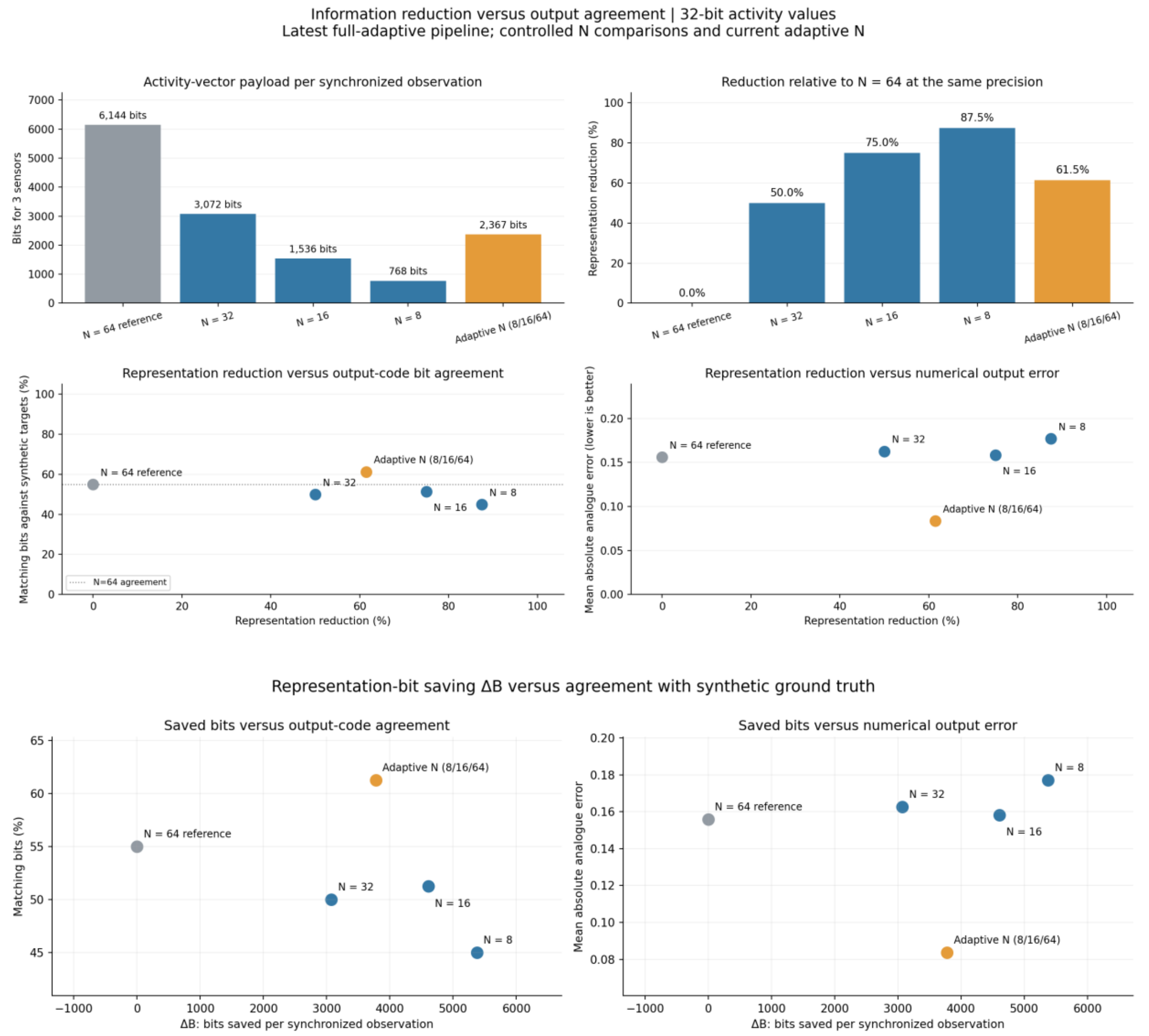

**Figure 17. Trade-off between sensor-information reduction and ALC output agreement. (a) Retained activity-vector payload, percentage reduction, output-code bit agreement, and mean absolute analogue error for fixed and scenario-adaptive activity-vector sizes; and (b) corresponding relationships between saved representation bits, bit-wise agreement, and numerical output error, demonstrating a 61.5% reduction in retained representation with improved synthetic target-code agreement for the adaptive configuration.**

# Challenges & Outlook

The findings of this analytical study highlight the possibility of shifting adaptability from the network level toward the fundamental processing and learning mechanisms of neuromorphic hardware. However, translating these theoretical models into practical circuits presents several challenges. The most critical is introducing scenario complexity as a continuously varying control parameter that can dynamically regulate information compression, reliability thresholds, pulse encoding, and particularly the nonlinear learning and forgetting dynamics of memristive devices. Achieving this coordination without introducing excessive computational or hardware overhead requires dedicated investigation. Equally important are identifying physical memory devices with suitable adaptive characteristics, managing analogue-circuit non-idealities, and developing compact spike-pattern decoders capable of translating limited neuronal activity into reliable actuator commands. Table 2 summarizes these four challenges and the potential implementation directions supported by recent developments.

While these developments provide possible implementation pathways, the critical question remains whether the benefits of adaptability can outweigh the additional hardware required to realize it. In particular, the ability to continuously estimate scenario complexity and use it to modulate the nonlinear dynamics of physical memory devices represents a central research challenge. Addressing this, together with device variability and efficient output decoding, will be essential for translating the proposed analytical framework into practical hardware.

**Table 2. Major challenges and potential implementation pathways for scenario-adaptive, reliability-aware neuromorphic sensor fusion.**

| Critical challenges | Possible implementation pathways | Key limitations and supporting references |
|---|---|---|
| 1. Continuous scenario-complexity estimation and hardware-level adaptation | Developing lightweight mixed-signal complexity estimators using activity counters, temporal monitors, and cross-sensor consistency circuits. The continuously estimated complexity could control programmable encoding thresholds, pulse generators, neuronal biasing, and memristive programming conditions through a shared adaptive controller. | The main difficulty is translating a continuously changing complexity value into coordinated control of multiple nonlinear processing stages, particularly the learning and forgetting characteristics of memristors, without excessive control overhead, calibration requirements, or instability. Memristor-based adaptive perception [43] demonstrates stimulus-dependent adaptation, while the Speck neuromorphic chip [44] demonstrates efficient event-driven preprocessing. Neither establishes the complete proposed complexity-controlled learning pipeline. |
| 2. Physical memory devices with adaptable learning and forgetting dynamics | Exploring volatile or hybrid volatile/nonvolatile memristors exhibiting tunable conductance updates, nonlinear switching, and multiple relaxation timescales. Pulse amplitude, duration, repetition, and history could be exploited to adjust learning rates and forgetting behavior. | A major challenge is finding devices whose plasticity and retention can be independently and reproducibly regulated according to the required scenario-dependent functions. Diffusive memristors [45] demonstrate intrinsic short- and long-term plasticity, while adaptive memristive perception [43] demonstrates configurable adaptation to sensory stimuli. |
| 3. Variability, noise, and reliability of analogue hardware | Developing variability-aware compact models, closed-loop programming, adaptive reference thresholds, device calibration, and circuit-level compensation mechanisms. Algorithm–hardware co-design could also exploit device stochasticity where beneficial. | Device-to-device variability, cycle-to-cycle switching fluctuations, conductance drift, temperature dependence, and analogue noise may disturb learning, forgetting, and neuronal firing patterns. Hardware-based Bayesian neural networks [46] demonstrate how memristive uncertainty can be incorporated into computation, while a recent review [47] discusses the broader hardware challenges. |
| 4. Compact and low-latency spike-pattern decoding for actuator control | Investigating temporal coincidence detectors, spike-count-based logic, time-to-first-spike decoding, compact lookup tables, and event-driven finite-state machines. Early decision mechanisms could generate actuator commands once sufficient evidence is available, without processing unnecessarily long spike sequences. | The key challenge is translating limited or sparse neuronal firing patterns into accurate control commands without complex decoders, long observation windows, or excessive power consumption. A dynamic SNN decoder [48] achieved 94.37% CIFAR-10 accuracy using an average of 1.69 time steps, corresponding to a 2.36× acceleration over its four-step baseline. A hardware time-to-first-spike implementation [49] demonstrated 196 ns average inference latency and 4.74 pJ per inference per neuron. These results motivate compact decoding, although reliable actuator-control validation remains necessary. |

Nevertheless, the present study provides a foundation for investigating ***how much information-processing and decision-making capability can be extracted from relatively simple neuron circuits by exploiting their tunable parameters and, eventually, the intrinsic nonlinear dynamics of physical memory devices.*** Future investigations should therefore focus on coordinated parameter optimization, experimental memristor characterization, circuit-level implementation, and validation against actual actuator-control decisions. Ultimately, these developments may contribute toward compact, energy-efficient, and fault-tolerant neuromorphic systems capable of continuously learning and responding at the sensor nodes, with little or no dependence on centralized computing resources.